%% file: monophoton_paper2017.tex
\pdfoutput=1 
\documentclass[cernpreprint,texlive=2016,,texmf,txfonts,UKenglish]{atlasdoc} 

\usepackage{atlaspackage}
\usepackage{atlasbiblatex}
\usepackage{atlasphysics}
\usepackage{atlasjournal}
\usepackage{atlasparticle}
\usepackage{atlasunit}
\usepackage{atlasxref}

\graphicspath{{logos/}{figures/}}

\usepackage{monophoton_paper2017-defs}

\input{atlas-preprint-cover}

\AtlasTitle{Search for dark matter at $\sqrt{s}=13\TeV$ in final states containing an energetic photon and large missing transverse momentum with the ATLAS detector}
\author{The ATLAS Collaboration}

\hypersetup{pdftitle={ATLAS draft},pdfauthor={The ATLAS Collaboration}}

\begin{document}
\maketitle

\tableofcontents
\clearpage

\section{Introduction}
\label{sec:intro}
\input{Introduction}

\section{The ATLAS detector}
\label{sec:detector}
\input{Detector}

\section{Monte Carlo simulation samples}
\label{sec:simsample}
\input{Mcarlo-samples}

\section{Event reconstruction}
\label{sec:Objects}
\input{Objects}

\section{Event selection}
\label{sec:Selection}
\input{Event-selection}

\section{ Background estimation}
\label{sec:BGEstimation}
\input{Background.tex}

\section{Results}
\label{sec:result}
\input{Results.tex}

\section{ Systematic uncertainties} 
\label{sec:syst}
\input{Systematics.tex}

\section{Interpretation of results}
\label{sec:interpretation}

\input{Interpretation.tex}

\FloatBarrier

\section{Conclusion}
\label{sec:conclusion}
\input{Conclusion}

\section*{Acknowledgements}
\input{Acknowledgements.tex}

\printbibliography
\clearpage
\newpage
\input{atlas_authlist}

\end{document}

%% file: atlas-preprint-cover.tex
\graphicspath{{logos/}}

\AtlasTitle{A standalone cover for CERN preprints}

\PreprintIdNumber{CERN-EP-2017-044}

\AtlasJournalRef{EPJC (2017) 77:393}
\AtlasDOI{10.1140/epjc/s10052-017-4965-8}

\AtlasAbstract{ %
\input{Abstract}
}

%% file: Abstract.tex
Results of a search for  physics beyond the Standard Model in events containing an energetic photon and large missing transverse momentum with the ATLAS detector at the Large Hadron Collider are reported. 
   As the number of events observed in data, corresponding to an integrated luminosity of  36.1 \ifb~
of proton--proton collisions at a centre-of-mass energy of $13\TeV$, is in agreement with the Standard Model expectations, 
model-independent  limits are set on the fiducial cross section for the production of events in this final state. 
Exclusion limits are also placed in models where dark-matter candidates are pair-produced. 
 For dark-matter production via an axial-vector or a vector mediator in the $s$-channel, this search excludes  
 mediator masses below  $750$--$1200\GeV$ for  dark-matter candidate masses below $230$--$480\GeV$ at 95\% confidence level, depending on the couplings.
 In an effective theory of dark-matter production, the limits restrict the value of the suppression scale $M_{*}$  to be above $790\GeV$ at 95\% confidence level.
A limit is also reported on the production of a high-mass scalar resonance by processes beyond the Standard Model, 
in which the resonance decays to $Z\gamma$ and the $Z$ boson subsequently decays into neutrinos.

%% file: Introduction.tex
Multiple theories of physics beyond the Standard Model (BSM) 
predict a high production rate of events containing a photon with a  high transverse energy ($\phPT$) and 
large missing transverse momentum (\metv, with magnitude \met) 
referred to as $\gamma+\missET$ events, in $pp$ collisions.
The search for  energetic $\gamma+\missET$ events, 
whose rates have a low expected contribution from Standard Model (SM) processes,
can thus provide sensitivity to new physics models~\cite{Abbiendi:2000hh,Achard:2003tx,Abdallah:2003np, Aaltonen:2008hh,Abazov:2008kp}, 
also related to dark matter (DM).
Although the existence of  DM is well established~\cite{Bertone:2004pz}, its nature is yet unknown.
 One candidate is a weakly interacting massive particle (WIMP, also denoted by $\chi$) that 
interacts  with SM particles with a strength similar to the weak interaction. If WIMPs interact with quarks via a 
mediator particle, pairs of WIMPs are produced in $pp$ collisions at sufficiently high energy. 
The $\chi\bar{\chi}$ pair is  invisible to the detector, but the radiation of an initial-state photon in $q\bar{q}\rightarrow\chi\bar{\chi}$ interactions~\cite{Goodman:2010ku} 
can produce detectable $\gamma+\missET$ events. 

\begin{figure}[htbp]
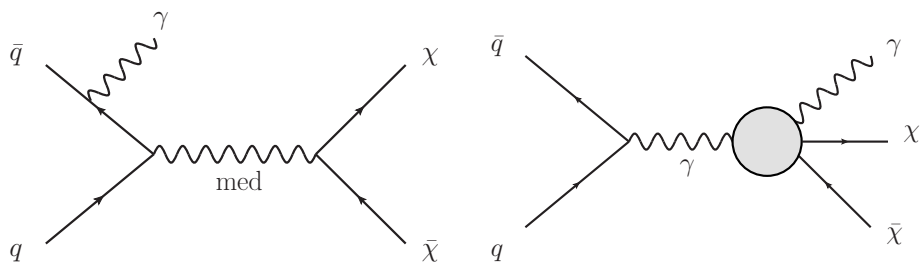
\centering
 \includegraphics[width=0.35\linewidth]{{{figures/fig_01a}}}
 \hspace{5.00mm}
 \includegraphics[width=0.35\linewidth]{{{figures/fig_01b}}}
 \caption{Pair production of dark-matter particles ($\chi\bar{\chi}$) in association with a photon via an explicit $s$-channel mediator (left), or via an effective $\gamma\gamma\chi\bar{\chi}$ vertex (right). 
 }
 \label{fig:FeynmanDiags}
\end{figure}

Effective field theories (EFT) with various forms of interaction between the WIMPs and the SM particles are a powerful  model-independent approach for the interpretation of DM production 
in $pp$ collisions~\cite{Goodman:2010ku}. However, the typical momentum transfer in $pp$ collisions at the LHC
can often exceed the cut-off scale below which the EFT approximation is valid.
Simplified models that involve the explicit production of the intermediate state, as shown in Figure~\ref{fig:FeynmanDiags} (left), avoid this limitation. 
This paper focuses on simplified models assuming Dirac-fermion DM candidates produced via an $s$-channel mediator with vector or axial-vector 
interactions~\cite{Abdallah, Buchmueller, dmforum}. There are  five free parameters in this model:
the WIMP mass $m_\chi$, the mediator mass $m_{\textrm{med}}$, the width of the mediator $\Gamma_{\textrm{med}}$, 
 the coupling $g_q$ of the mediator to quarks, and the coupling $g_{\chi}$ of the mediator to the dark-matter particle.
In the limit of a large mediator mass, these simplified models map onto the EFT operators, with the suppression scale\footnote{The suppression scale, also referred to as $\Lambda$, is the effective mass scale of particles that are integrated out in an EFT. The non-renormalisable operators are suppressed by powers of 1/$\Mstar$.} 
$\Mstar$ linked to $m_{\textrm{med}}$ by the relation ${\Mstar=m_{\textrm{med}}/\sqrt{g_qg_\chi}}$~\cite{Busoni}.  

The paper also considers a specific dimension-7 EFT operator with direct couplings between DM and electroweak (EW) bosons, 
for which there is neither a corresponding simplified model nor a simplified model yielding similar kinematic distributions implemented in 
an event generator~\cite{Crivellin,dmforum}.
The process describing a contact interaction of type $\gamma\gamma\chi\bar{\chi}$ is shown in Figure~\ref{fig:FeynmanDiags} (right).
In this model, DM production proceeds via 
$q \bar{q} \rightarrow \gamma \rightarrow \gamma\chi\bar{\chi}$, generating an energetic photon  without requiring initial-state radiation.  
There are four free parameters in this model: the EW coupling strengths $k_1$ and $k_2$ (which respectively control the strength of the coupling 
to the SM U(1) and SU(2) gauge sectors), $m_{\chi}$, and the suppression scale $\Mstar$.

Many BSM models~\cite{BSMRes, BSMRes2} introduce new bosons through either an extension of the Higgs sector or additional gauge fields.
In some of those, the bosons are predicted to decay into electroweak gauge bosons: 
the analysis presented here also searches for such a resonance decaying into $Z\gamma$, which would lead to an excess
 of energetic $\gamma$+\met\ events when the $Z$ boson subsequently decays to neutrinos. 

The ATLAS and CMS collaborations
 have reported limits in various models based on searches for an excess of 
 $\gamma+\missET$ events using $pp$ collisions at centre-of-mass energies of $\sqrt{s}=7$~and $8 \TeV$ (LHC Run 1) and with  the first LHC Run-2 data collected in 2015 
 at a centre-of-mass energy of $13\TeV$~\cite{Aad:2012fw, PRD2015, Monophoton2015, Chatrchyan:2012tea, CMS_8TeV}. 
 A $\chi\bar{\chi}$ pair can also be produced in association with other objects  leading to different $X$+\met signatures, 
 where $X$ can be  a  jet, a $W$ boson, a $Z$ boson or a Higgs boson.
 DM searches are hence performed in a variety of complementary final states \cite{Aaboud:2016tnv, Aaboud:2016qgg, Aaboud:2016obm, Sirunyan:2017hci, Sirunyan:2017hnk,}.
 The $\gamma+\missET$ final state has the advantage of a clean signature 
providing a good complementarity with respect to the other  $X$+\met 
processes. Moreover it also offers the unique possibility to probe for 
DM models in which the photon does not come from initial-state radiation.
 This paper reports the results of a search for dark matter and  for a BSM $Z\gamma$  resonance 
 in $\gamma+\missET$ events in $pp$ collisions at a centre-of-mass energy of $\sqrt{s}=13\TeV$ using the Run-2 data collected in 2015 and 2016, 
 corresponding to an integrated luminosity of 36.1 \ifb.
As described in Section~\ref{sec:Selection}, this search follows a strategy similar to that implemented  
in Ref.~\cite{Monophoton2015}, but with multiple signal regions optimised to take advantage of the tenfold increase in integrated luminosity. 

The paper is organised as follows.
A brief description of the ATLAS detector is given in Section \ref{sec:detector}. 
The signal and background Monte Carlo (MC) simulation samples used are described in Section \ref{sec:simsample}.
The reconstruction of physics objects is explained in Section \ref{sec:Objects}, and
the event selection is described in Section \ref{sec:Selection}.
Estimation of the SM backgrounds is outlined in Section \ref{sec:BGEstimation}. 
The results  are described in Section \ref{sec:result} and the systematic uncertainties are given in Section \ref{sec:syst}.
The interpretation of results in terms of  models of pair production of dark-matter candidates and of BSM production of a high-mass $Z\gamma$ resonance 
is described in  Section \ref{sec:interpretation}. 
A summary is given in Section \ref{sec:conclusion}.

%% file: Detector.tex
The ATLAS detector \cite{ATLAS} is a multipurpose particle
physics apparatus with a forward--backward symmetric
cylindrical geometry and near 4$\pi$ coverage in solid angle.\footnote{ATLAS uses a right-handed coordinate system with 
its origin at the nominal  interaction point (IP) in the centre of the detector and the $z$-axis along the beam pipe. The $x$-axis points from the IP to the centre of the LHC ring, and the $y$-axis points upward. Cylindrical coordinates ($r,\phi$) are used in the transverse plane, $\phi$ being the azimuthal angle around the beam pipe. The pseudorapidity is defined in terms of the polar $\theta$ angle as ${\eta = -\ln\left[\tan(\theta/2)\right]}$.}
The inner tracking detector (ID), covering the
pseudorapidity range $|\eta| <$ 2.5, consists of a silicon
pixel detector including the insertable B-layer~\cite{ATLAS-TDR-19, ATLAS-TDR-19_a}, which was added around a new, smaller-radius beam-pipe before the start of Run 2; a silicon microstrip detector;
 and, for $|\eta| <$ 2.0, a straw-tube transition radiation tracker (TRT). 
The ID is surrounded by a thin superconducting solenoid which provides
a 2 T magnetic field.
 A high-granularity lead/liquid-argon sampling electromagnetic
 calorimeter (EM) covers the region $|\eta|<$~3.2. It 
 is segmented longitudinally in shower depth.
The first layer has a high granularity in the $\eta$ direction in order to provide an efficient discrimination between single-photon showers and two overlapping photons originating from a $\pi^0$ decay.
The second layer is where most of the energy, deposited in the calorimeter by electron- or photon-initiated electromagnetic showers, is collected.
Significant energy deposits can be left in the third layer by very high energy showers; this layer can also be used to correct for energy leakage beyond the 
electromagnetic calorimeter.
A steel/scintillator-tile hadronic calorimeter covers the range $|\eta| <$ 1.7, 
while the liquid-argon technology with either copper or tungsten as the absorber material is used
for the hadronic calorimeters in the end-cap region 1.5~$<|\eta|<$~3.2 and for electromagnetic and hadronic measurements in the forward region up to $|\eta|=$~4.9.
A muon spectrometer (MS) surrounds
the calorimeters. It consists of three large air-core superconducting toroidal magnet systems, precision tracking chambers providing 
accurate muon tracking out to $|\eta|$~=~2.7, and fast detectors for triggering in the region $|\eta| <$~2.4. 
A two-level trigger system is used to select events for offline analysis~\cite{Trigger_new}.

%% file: Mcarlo-samples.tex
Several simulated MC samples are used to estimate the signal acceptance, the detector efficiency and various SM
background contributions.
For all the DM samples considered here, the values of the free parameters were chosen following the recommendations given in Ref.~\cite{dmforum}.

Samples of DM production in simplified models are generated via an $s$-channel mediator with axial-vector interactions.
The program  \textsc{MG5}\_a\textsc{MC@NLO} v2.4.3~\cite{Alwall:2014hca} is used in conjunction with $\pythia$ v8.212~\cite{Sjostrand:2007gs} 
with the parameter values set according to the ATLAS tune A14~\cite{ATL-PHYS-PUB-2014-021}. 
The parton distribution function (PDF) set used is NNPDF3.0 at next-to-leading order (NLO)~\cite{Ball:2014uwa} with $\alpha_{\mathrm{s}}$ = 0.118.
The $g_q$ coupling is set to be universal in quark flavour and equal to 0.25,
the $g_{\chi}$ coupling is set to 1.0, and $\Gamma_{\textrm{med}}$ is computed as the minimum width allowed given the couplings and masses.
As shown in Ref.~\cite{dmforum}, $\Gamma_{\textrm{med}}/m_{\textrm{med}}$ varies between 2\% and 6\% for the values probed here. 
Different choices of the couplings and a model with a vector mediator are also considered, as described in Section~\ref{sec:interpretation}.
 The generation was updated with respect to the 2015 data analysis~\cite{Monophoton2015} by using the DMsimp~\cite{Backovic:2015soa} implementation 
 of the model at NLO. 
Events are generated with parameters spanning a grid of points in the $m_{\chi}$--$m_{\textrm{med}}$ plane.  

For DM samples corresponding to an EFT model involving dimension-7 operators with a contact interaction of type $\gamma\gamma\chi\bar{\chi}$,
the parameters which only influence the cross section are set to $k_1=k_2=1.0$ and $M_*=3.0\TeV$~\cite{dmforum}. A scan over a range of values of $m_{\chi}$ is performed. 
Events are generated with \textsc{MG5}\_a\textsc{MC@NLO} v2.2.3 
and the PDF set NNPDF3.0 at leading order (LO) with $\alpha_s$ =0.130, 
in conjunction with $\pythia$~v8.186, using the ATLAS tune A14.

For DM signal generation in both the simplified and EFT models, a photon with at least $\phPT = 130\GeV$  is required at the matrix-element level in \textsc{MG5}\_a\textsc{MC@NLO}.

The samples used in the search for a BSM high-mass scalar resonance decaying to  $Z\gamma$  are generated using \textsc{Powheg-Box} v1~\cite{Alioli:2010xd}, with the CT10 PDF set~\cite{Lai:2010vv} and $\pythia$~v8.210 for the showering with the AZNLO tune~\cite{Aad:2014xaa} based on the CTEQ6L1 PDF set~\cite{Pumplin:2002vw}. 
The simulated heavy scalar resonance $X$ of very narrow width (4 \MeV), with masses in the range 2 to  $5\TeV$, is produced  through gluon--gluon fusion and then assumed to decay exclusively to $Z\gamma$.

For all the signal samples described above, the EvtGen v1.2.0 program~\cite{EvtGen} is used for properties of the bottom and charm hadron decays.

For $W\gamma$ and $Z\gamma$ backgrounds, events containing a charged lepton ($e$, $\mu$ or $\tau$) and a neutrino, 
a pair of neutrinos ($\nu\nu$) or a pair of charged leptons ($\ell\ell$) together with a photon and associated jets are simulated using the $\sherpa$ v2.1.1 event generator~\cite{Gleisberg:2008ta}. 
The matrix elements including all diagrams with three electroweak couplings are calculated 
with up to three partons at LO and merged with $\sherpa$ parton showers~\cite{Schumann:2007mg} using the ME+PS@LO prescription 
\cite{Hoeche:2009rj}. The CT10 PDF set is used in conjunction with a dedicated parton shower tuning developed by the 
$\sherpa$ authors. 
For $Z$ events with the $Z$ boson decaying to a $\ell\ell$ pair  a requirement on the dilepton invariant mass of $m_{\ell\ell} > 10 \GeV$ 
is applied at event generator level.

Events containing a photon with associated jets are also simulated using $\sherpa$ v2.1.1~\cite{Gleisberg:2008ta}, generated in several bins of $\phPT$  
with lower edges ranging from $35\GeV$ to $4\TeV$. The matrix elements are calculated at LO with up to three or four partons 
 and merged with $\sherpa$ parton showers using the ME+PS@LO prescription. The CT10 PDF set is used in conjunction with the  
dedicated parton shower tuning.

For $W/Z$+jets backgrounds, events containing $W$ or $Z$ bosons with associated jets are simulated using $\sherpa$ v2.2.0.
The matrix elements are calculated for up  to four partons at LO and two partons at NLO 
using the Comix~\cite{Gleisberg:2008fv} and OpenLoops~\cite{Cascioli:2011va} matrix-element generators and merged with $\sherpa$ parton showers using the ME+PS@NLO prescription~\cite{Hoeche:2012yf}. 
The NNPDF3.0 PDF set at next-to-next-to-leading order (NNLO) is used. As in the case of the $\gamma$+jets samples, the dedicated parton shower tuning is used. The $W/Z$+jets events are normalised to the NNLO inclusive cross sections~\cite{ATL-PHYS-PUB-2016-003}.

Table~\ref{tab:MCSamples}  summarises  the choices made in the generation of MC samples used in the
analysis. 

\begin{table}[!h]
\small
       \begin{center}
       \caption{Details of the generation of the signal samples and of the SM background processes considered in the analysis.}
       \label{tab:MCSamples}
       \begin{tabular}{ l l l l l l r }
          \hline
       \hline
        Process & Event generators used & PDF sets & Order &  Requirements\\
        \hline
                            &                                                                        &                                                      &                                              &\\[-0.5em] 
         DMsimp model & \textsc{MG5}\_a\textsc{MC@NLO} v2.4.3  & NNPDF30\_nlo\_as\_0118 & NLO &  $\phPT > 130\GeV$\\
                                  & + $\pythia$~v8.212                                                   &              
                                  &         & \\
                            &                                                                        &                                                      &                                              &\\ [-0.5em]                                 
         EFT model & \textsc{MG5}\_a\textsc{MC@NLO} v2.2.3   & NNPDF30\_lo\_as\_0130 & LO &  $\phPT > 130\GeV$\\
                                  & + $\pythia$~v8.186                                                   &                
                                  &         &\\
                             &                                                                        &                                                      &                                              &\\[-0.5em] 
         BSM resonance & \textsc{Powheg-Box} v1 & CT10 & NLO   &  $--$\\
                                                        &+ $\pythia$~v8.210               &         &       & \\
                            &                                                                        &                                                      &                                              &\\[-0.5em]                                                         
        $W/Z\gamma$ & $\sherpa$ v2.1.1 & CT10 & LO   &for $Z$:  $m_{\ell\ell} > 10 \GeV$\\
                            &                                                                        &                                                      &                                              &\\ [-0.5em] 
        $\gammajets$ & $\sherpa$ v2.1.1& CT10 & LO    & $--$\\
                            &                                                                        &                                                      &                                              &\\[-0.5em]  
        $W/Z$+jets & $\sherpa$ v2.2.0 &  NNPDF30\_nnlo & LO/NLO    & $--$\\
                             &                                                                        &                                                      &                                              &\\[-0.5em]  
                              &                                                                        &                                                      &                                              &\\[-0.5em] 
       \hline
       \hline
       \end{tabular}
       \end{center}
\end{table}

Multiple $pp$  interactions in the same or neighbouring bunch crossings (referred to as  pile-up)
superimposed on the hard physics process  are simulated with the minimum-bias  processes of $\pythia$~v8.186 using the A2 
tune~\cite{ATL-PHYS-PUB-2011-014} and the MSTW2008LO PDF set~\cite{Martin:2009iq}.
Simulated events are reweighted so that the distribution of the expected number of collisions per bunch crossing, $\langle \mu \rangle$, matches the one observed in 
data, which has a mean value of 13.7 (24.2) in  2015 (2016) data.

All generated event samples are processed with a full ATLAS detector simulation~\cite{:2010wqa} based on
\textsc{Geant4}~\cite{Agostinelli:2002hh}.
The simulated events are reconstructed and analysed with the same analysis chain as used for the data,
utilising  the same trigger and event selection criteria discussed in Section \ref{sec:Selection}.

%% file: Objects.tex
Photons are reconstructed from clusters of energy deposits in the electromagnetic calorimeter measured in projective towers.
Clusters without matching tracks are classified as unconverted photon candidates.
A photon candidate containing clusters that can be matched to tracks is considered as a converted photon candidate~\cite{ATL-PHYS-PUB-2011-007}.
The photon energy is corrected by applying the energy scales measured with
$Z \rightarrow e^+ e^-$ decays~\cite{PERF-2013-05}.
The trajectory of the photon is reconstructed using the longitudinal (shower depth) segmentation of the calorimeters and a constraint from the average collision point of the proton beams.
For converted photons, the position of the conversion vertex is also used if tracks
from the conversion have hits in the silicon detectors. 
Identification requirements are applied in order to reduce the contamination from $\pi^0$ or other neutral hadrons 
decaying to two photons. The photon identification is based on the profile of the energy deposits in the first  and second layers of the
 electromagnetic calorimeter. 
 Candidate photons are required to have $\phPT > 10\GeV$, to satisfy the "loose" identification criteria defined in Ref.~\cite{PhId} 
  and to be within $\abseta < 2.37$. 
  Photons used in the event selection must additionally satisfy the "tight" identification criteria~\cite{PhId}, have $|\eta|<1.37$ or $1.52<|\eta|<2.37$ and be isolated 
by requiring the energy in the calorimeters in a cone of size $\Delta R = \sqrt{(\Delta \eta)^2 + (\Delta \phi)^2} = 0.4$ around the 
cluster barycentre, excluding the energy associated with the photon cluster, to be less than $2.45\GeV + 0.022\times\phPT$.
This cone energy is corrected for the leakage of the photon energy from the central core and for the
effects of pile-up~\cite{PERF-2013-05}.
In addition, the scalar sum of the $\pt$ of non-conversion tracks in a cone of size $\Delta R = 0.2$ around the cluster barycentre is required to be less than $0.05\times\phPT$.

Electrons are reconstructed from clusters in the electromagnetic calorimeter which are matched to a track in the ID.
The criteria for their identification, and the calibration steps, are similar to those used for photons.
Electron candidates must fulfil the "medium" identification requirement of Ref.~\cite{PERF-2013-05}.
Muons are identified either as a combined track in the MS and ID systems,
or as an ID track that, once extrapolated to the MS, is associated with at least one track segment in the MS.
Extrapolated muons are also considered; they are reconstructed from an 
 MS track which is required to be compatible with originating from the nominal  interaction point.
Muon candidates must pass the "medium" identification requirement~\cite{Muons13TeV}.
The significance of the transverse impact parameter, defined as the transverse impact parameter $d_0$ divided by its estimated
uncertainty, $\sigma_{d_0}$, of tracks with respect to the beam line 
 is required to satisfy $|d_0|/\sigma_{d_0} <$ 5.0 for 
electrons and $|d_0|/\sigma_{d_0} <$ 3.0 for muons. The longitudinal impact parameter $z_0$ must satisfy ${|z_0 \sin\theta| < 0.5 ~{\mathrm{mm}}}$
for both the electrons and muons.
Electrons are required to have $\pt > 7\GeV$ and $\abseta < 2.47$, while muons are 
required to have $\pt > 6\GeV$ and $\abseta  < 2.7$.

Jets are reconstructed with the anti-$k_{t}$ algorithm~\cite{Cacciari:2008gp} with a radius parameter $R = $ 0.4 
from clusters of energy deposits at the electromagnetic scale in the calorimeters~\cite{topoclusters}.
A correction used to calibrate  the jet energy  to the scale of its constituent particles~\cite{PERF-2016-04, SingleHad}
is then applied.
Jets are also corrected for contributions from pile-up interactions
and a residual correction derived in situ is applied as the last step to jets reconstructed in data~\cite{PERF-2016-04} .
Candidate jets are required to have $\pt > 20\GeV$.
In order to suppress pile-up jets, which are mainly at low \pt, a jet vertex tagger~\cite{JVT}, based on tracking and vertexing information, 
is applied for jets with \pt\ $< 60\GeV$~  and $\abseta<2.4$. 
Jets used in the event selection are required to have \pt\ $>30\GeV$ and $|\eta| < 4.5$.
The $\tau$ leptons decaying to hadrons and $\nu_{\tau}$ are considered as jets as in previous analyses~\cite{PRD2015,Monophoton2015}.

The removal of overlapping candidate objects is performed in the following order.
If any selected electron shares its ID  track with a selected muon, the electron is removed and the muon is kept, in order to 
remove electron candidates originating from muon bremsstrahlung followed by photon conversion.
If an electron lies a distance $\Delta R < 0.2$ of a candidate jet, the jet is removed
 from the event, while if an electron lies a distance $0.2 < \Delta R < 0.4$ of a  jet, the electron is removed.
Muons lying a distance $\Delta R < 0.4$ with respect to the remaining candidate jets
are removed, except if the number of tracks with $\pt > 0.5\GeV$  associated with the jet is less
than three. 
In the latter case, the muon is kept and the jet is discarded. 
Finally, if a jet lies a distance $\Delta R < 0.4$ of a candidate photon, the jet is removed.

The missing transverse momentum vector \metv~is obtained from the negative vector sum of the momenta of the 
candidate physics objects, selected as described above.
Calorimeter energy deposits and tracks are matched with candidate high-$\pt$ objects in a specific order: 
electrons with \pt$ > 7\GeV$, photons with $\phPT > 10\GeV$, muons with \pt$ > 6\GeV$ and jets with \pt$ > 20\GeV$ \cite{Aad:2012re}.
Tracks from the primary vertex\footnote{The primary vertex is defined as  the vertex with the highest sum of the squared transverse momenta of its associated tracks. 
It is reconstructed from at least two associated good-quality tracks with $\pt > 0.4\GeV$~\cite{ATL-PHYS-PUB-2015-026}.}
  not associated with any such objects  are also taken into account in the \met\ 
reconstruction ("soft term") \cite{MET_Paper_8TeV}.

Corrections are applied to the objects in the simulated samples to account for differences compared to data in object reconstruction, identification and isolation efficiencies for both 
the leptons and photons. For the photons, the efficiency corrections depend on whether or not they are converted, and on their $\phPT$ and $\eta$; 
for the photons used in this analysis they are generally of the order of $1\%$.

%% file: Event-selection.tex
The data were collected in $pp$ collisions at $\sqrt{s}=13 \TeV$  during 2015 and 2016. 
The events for the analysis were recorded using a trigger 
requiring at least one photon candidate above a $\phPT$ threshold of $140\GeV$ to pass "loose" identification requirements,
which are based on the shower shapes in the EM calorimeter 
as well as on the energy leaking into the hadronic calorimeter~\cite{PhIsol}. 

For events in the signal regions defined below, the efficiency of the trigger is more than 98.5\%.
The 1\% difference in the efficiency between data and MC simulation is treated as a systematic uncertainty.
Only data satisfying beam, detector and data-quality criteria are considered.
The data used for the analysis correspond to an integrated luminosity of 36.1 \ifb. 
The uncertainty in the integrated luminosity is $\pm$3.2\%.
It is derived following a methodology similar to that detailed in Ref.~\cite{lumi2016}, from a preliminary calibration of the 
luminosity scale using $x$--$y$ beam-separation scans performed in August 2015 and May 2016. 

Events are removed if they contain a bad-quality photon or jet~\cite{ATL-PHYS-PUB-2011-007, ATLAS-CONF-2012-020}, arising from instrumental problems or non-collision background.
Events are required to have a reconstructed primary vertex, as defined in Section~\ref{sec:Objects}.

Events in the signal regions (SRs) are required to have the leading photon satisfying the criteria defined in Section~\ref{sec:Objects} and
having  $\phPT>150 \GeV$, which is well above the thresholds used for the MC event generation and for the data-collection trigger.
The |$z$| position, defined as the longitudinal separation between the beamspot position and the intersection of the extrapolated photon trajectory with the beam-line, is required to be smaller than 0.25~m.
This criterion provides a powerful rejection against the muons from beam background~\cite{Monophoton2015}, 
which can leave significant energy deposits in the calorimeters
and hence lead to reconstructed fake photons that do not point back to the primary vertex.
 It is required that the photon and \metv\ do not overlap in the azimuthal plane: $\Delta\phi(\gamma,~$\metv$) > 0.4$.
 To further suppress the background events where the $\met$ is fake,  e.g.\ arising from poorly reconstructed objects, a requirement on the ratio $\met/\sqrt{\Sigma \ET} > 8.5~{\textrm{GeV}^{1/2}}$ 
 is added,\footnote{This ratio simulates the $\MET$ significance because the $\MET$ resolution due to purely calorimetric measurements scales approximately as $\sqrt{\Sigma \ET}$.} 
 where $\Sigma \ET$ is calculated as the scalar sum of all $\pt$ from the objects and the tracks contributing to the $\MET$ reconstruction described in Section~\ref{sec:Objects}.  
 This requirement mainly rejects the $\gamma$+jets background events.
Events with more than one jet or with a jet with $\Delta\phi(\jet,~$\metv$)<0.4$ are rejected (jet veto), the latter to remove events where $\met$ originates from jet mismeasurement.
The remaining events with one jet are retained to increase the signal acceptance and reduce systematic uncertainties related to the modelling of initial-state radiation.
Events are required to have no electrons  or muons passing the requirements for $e/\mu$ candidates described in Section \ref{sec:Objects}.
This lepton veto mainly rejects $W/Z$ events with charged leptons in the final state.

As the production of a pair of dark-matter candidates or of a high-mass BSM $Z(\rightarrow \nu\nu)\gamma$ resonance are both expected to lead to events with large $\MET$, 
 five SRs are defined with different $\MET$ requirements: three inclusive (SRI1, SRI2 and SRI3) with increasing $\MET$ thresholds 
and two exclusive (SRE1 and SRE2)  with $\MET$ in two different ranges.
Table~\ref{tab:SR_EvSel} shows the criteria for selecting events in the SRs and  the number of events selected in data. 
The fraction of events in which the selected photon is unconverted  is about $70\%$ for all regions. 
The fraction of selected events with no jets increases in the regions with lower $\met$ thresholds and ranges from about $50\%$ to about $70\%$.

\begin{table}[!htbp]
\begin {center}
\caption{Criteria for selecting events in the SRs and the numbers of events selected in data.}
\label{tab:SR_EvSel}
\begin{tabular}{c D{,}{\pm}{4,4} D{,}{\pm}{4,4} D{,}{\pm}{4,4} D{,}{\pm}{4,4} D{,}{\pm}{4,4} }
\toprule
\toprule
\multicolumn{1}{l}{Event cleaning} & \multicolumn{5}{c}{Quality and Primary vertex}  \\
  &                        &                                                                        &                                                      &                                              &\\[-0.8em] 
\multicolumn{1}{l}{Leading photon} &\multicolumn{5}{c}{$\phPT > 150 \GeV$, $|\eta|<1.37$ or $1.52<|\eta|<2.37$,}  \\
\multicolumn{1}{c}{} &\multicolumn{5}{c}{tight, isolated, $|z| < 0.25~{\textrm m}$,}  \\
\multicolumn{1}{c}{} &\multicolumn{5}{c}{$\Delta\phi(\gamma,~$\metv$) > 0.4$} \\
  &                        &                                                                        &                                                      &                                              &\\[-0.8em] 
\multicolumn{1}{l}{$\met/\sqrt{\Sigma \ET}$} &\multicolumn{5}{c}{$ > 8.5 ~{\textrm{GeV}^{1/2}}$} \\
 &                          &                                                                        &                                                      &                                             &\\[-0.8em] 
\multicolumn{1}{l}{Jets} &\multicolumn{5}{c}{0 or 1  with  \pt\ $>30\GeV$, $|\eta| < 4.5$ and $\Delta\phi(\jet,~$\metv$)>0.4$} \\
 &                          &                                                                        &                                                      &                                             &\\[-0.8em] 
\multicolumn{1}{l}{Lepton} &\multicolumn{5}{c}{veto on $e$ and $\mu$} \\
\midrule
& \multicolumn{1}{c}{ SRI1}& \multicolumn{1}{c}{ SRI2}& \multicolumn{1}{c}{ SRI3}&
\multicolumn{1}{c}{ SRE1}& \multicolumn{1}{c}{ SRE2}\\
\multicolumn{1}{l}{ $\MET~[{\textrm{GeV}}]$ }& \multicolumn{1}{c}{$ > 150$ } & \multicolumn{1}{c}{$ > 225$ 
} &  \multicolumn{1}{c}{ $ > 300$ } &  \multicolumn{1}{c}{$150$--$225$} &  \multicolumn{1}{c}{$225$--$300$} \\
 &                          &                                                                        &                                                      &                                             &\\[-0.8em] 
\midrule
\multicolumn{1}{l}{Selected events in data} & \multicolumn{1}{c}{2400} & \multicolumn{1}{c}{729} &  \multicolumn{1}{c}{236} &  \multicolumn{1}{c}{1671} &  
\multicolumn{1}{c}{493} \\

\multicolumn{1}{l}{Events with 0 jets} & \multicolumn{1}{c}{1559} & \multicolumn{1}{c}{379 } &  \multicolumn{1}{c}{116 } & 
 \multicolumn{1}{c}{1180 } &  \multicolumn{1}{c}{ 263 } \\

\bottomrule
\bottomrule
\end{tabular}
\end {center}
\end{table}

%% file: Background.tex
The SM background to the $\gamma+\missET$ final state is due to
events containing either a true photon or an object misidentified as a photon.
The background with a true photon is dominated by the electroweak production of  $Z(\rightarrow \nu \nu)\gamma$ events.
Secondary contributions come from $W(\rightarrow \ell\nu)\gamma$  and $Z(\rightarrow \ell\ell)\gamma$ production with unidentified electrons, muons or with  $\tau \rightarrow \textrm{hadrons}+\nu_{\tau}$ decays  
and from $\gamma$+jets events. The contribution from $t\bar{t}+\gamma$ is negligible thanks to the jet veto.
The contribution from  events where  a lepton or a jet is misidentified as a photon arises mainly from $W/Z$+jets production, 
with smaller contributions from diboson, multi-jet and  top-quark pair production.

All  significant background estimates are extrapolated from non-overlapping data samples.
A simultaneous fit in background-enriched control regions (CRs) is performed to obtain normalisation factors for the  $W\gamma$, $Z\gamma$ and $\gamma$+jets backgrounds  
(see Sections \ref{sec:bVgamma} and \ref{sec:bGammajet}), 
which are then used to estimate backgrounds in the SRs;  more details are given in Sections~\ref{inclusive_fit} and \ref{multiple_fit}.
The same normalisation factor is used for 
both $Z(\rightarrow \nu\nu)\gamma$ and $Z(\rightarrow \ell\ell)\gamma$ in SR events.   
The backgrounds due to photons from the misidentification of
electrons or jets in processes such as $W/Z$+jets, diboson and multi-jet events
(referred to as fake photons) are estimated using data-driven techniques (see Sections \ref{sec:bElefake} and \ref{sec:bJetfake}).

\subsection{ $W\gamma$ and $Z\gamma$ backgrounds}
\label{sec:bVgamma}
  
   For the estimation of the $W/Z\gamma$ background, 
three control regions are defined for each SR by selecting events with the same criteria used for the various SRs but inverting the lepton vetoes. 
  As the $\gamma$+jets background contribution is not significant in these leptonic CRs, the requirement on the ratio $\met/\sqrt{\Sigma \ET}$ is not applied.
  In the one-muon control region (1muCR) the $W\gamma$ contribution is enhanced by requiring the presence of a muon;
the 1muCR is sufficient to constrain the $W\gamma$ normalisation effectively without the need of a similar one-electron control region,
which would be contaminated by $\gamma$+jets  background.
  The two-lepton control regions enhance the $Z\gamma$ background by requiring the presence of a pair of muons (2muCR) or electrons (2eleCR).
  In each case, the CR lepton selection follows the same requirements as the SR lepton veto, with the addition that the leptons must be isolated with "loose" criteria~\cite{Muons13TeV} 
  based on information from the calorimeter and tracking systems.  
  In both 1muCR and 2muCR, the $\met$ value is computed disregarding 
muons, effectively treating them as non-interacting particles, in order to ensure that 
the $\met$ distributions in those CRs are similar to that in the SR.
  The same procedure is followed for electrons in 2eleCR.
  In both the $Z\gamma$-enriched control regions,  the dilepton invariant mass $m_{\ell \ell}$ is required to be greater than $20 \GeV$, and the invariant mass of the leptons and photon, $m_{\ell \ell \gamma}$, is required to be smaller than $1\TeV$ in order to avoid probing for potential BSM high-mass $Z\gamma$ resonances.
  The normalisation of the dominant $Z(\rightarrow \nu\nu)\gamma$ background source is largely constrained by the event yields in 2muCR and 2eleCR.
  The systematic uncertainty due to the extrapolation of the correction factors from CRs to SRs is taken into account (see Section~\ref{sec:syst}).

\subsection{$\gamma$+jets background}
\label{sec:bGammajet}

The $\gamma$+jets background in the SRs consists of events where the jet is poorly
reconstructed and partially lost, creating  fake $\MET$.
This background, which  increased in  2016 data  with respect to 2015 data because of the higher pile-up   
 conditions,   is suppressed by the large $\MET$ and jet--\metv\ azimuthal separation requirements and
 by the requirement $\met/\sqrt{\Sigma \ET} > 8.5~{\textrm {GeV}^{1/2}}$ described in  Section~\ref{sec:Selection}. 
 This last requirement reduces the contribution of $\gamma$+jets events in SRI1 to less than 10\% of the total background, 
 with a negligible effect on the acceptance for signal events.
 The fraction of $\gamma$+jets events decreases with $\met$ and becomes  
 less than $2\%$ of the total background in SRI3.
For the estimation of the residual $\gamma$+jets background, a specific control region (PhJetCR) enriched in $\gamma$+jets events is defined.
 It uses the same criteria as used for the SRs, but does not apply the requirement on the ratio $\met/\sqrt{\Sigma \ET}$, and requires  $85\GeV <  \met < 110\GeV$ and 
azimuthal separation between the photon and \metv,  $\Delta\phi(\gamma,~$\metv$)$, to be smaller than 3.0.
The latter selection minimises the contamination 
from signal events, which is estimated to be at most at the level of 1\%.
The PhJetCR  is the same for all SRs;
the systematic uncertainty due to the modelling of the $\gamma$+jets background, which affects its extrapolation from 
the low-$\met$ PhJetCR to the SRs  with larger $\met$, is taken into account (see Section~\ref{sec:syst}).

\subsection{ Fake photons from misidentified electrons}
\label{sec:bElefake}
Contributions from processes in which an electron is misidentified as a photon in the SRs are estimated by scaling yields from a sample of 
$e +\missET$ events by an electron-to-photon misidentification factor. This factor is measured with mutually exclusive samples of 
$e^+e^-$ and $\gamma+e$ events in data. To establish a pure sample of electrons, the $ee$ and the $e\gamma$ invariant masses  
($m_{ee}$ and $m_{e\gamma}$) are both required to be 
consistent with the $Z$ boson mass to within $10\GeV$, and the $\missET$ is required to be smaller than $40\GeV$. 
Furthermore, the sidebands to the $Z$ boson mass window are used to estimate and subtract possible contamination from misidentified jets in this sample.
The  misidentification factor, calculated as the
ratio of the number of $\gamma+e$  to  the number of $e^+e^-$  events, is parameterised as a function of $\pt$  and pseudorapidity and it 
varies between $0.59\%$ and  $2.5\%$.
Systematic uncertainties in the misidentification factors are evaluated by 
varying the sideband definition, comparing the results of the 
method (with or without using the sideband subtraction) with generator-level information about $Z(\rightarrow ee)$ MC events, 
and comparing the misidentification factors in $Z(\rightarrow ee)$ and $W(\rightarrow e\nu)$ MC events.
Background estimates are then also made for the four control regions, 1muCR, 2muCR, 2eleCR and PhJetCR,  by applying the electron-to-photon misidentification factor to events selected 
with the same criteria used for these regions but requiring an electron instead of a photon.
The estimated contribution from this background in the SRs and the associated uncertainty are reported in Section~\ref{sec:result}. 

\subsection{ Fake photons from misidentified jets}
\label{sec:bJetfake}
Background contributions from events in which a jet is misidentified as a photon are estimated
using a sideband counting method~\cite{PhIsol}.
This method relies on counting photon candidates in four regions of a
two-dimensional space, defined by the  isolation transverse energy and by
the quality of the identification criteria.
 A signal region (region A) is defined by photon candidates that are isolated with tight identification. 
Three background regions are defined, consisting of photon candidates which are tight and 
non-isolated (region B), non-tight and isolated (region C) or non-tight and non-isolated (region D). 
The method relies on the assumption that the isolation profile in the non-tight region is the same as that of the background in the tight region. 
This has been verified in MC samples by checking that the correlation factor, calculated as ($N_{\textrm A }* N_{\textrm D }/ N_{\textrm B} * N_{\textrm C}$) is compatible with unity within uncertainties.
The number of background candidates in the signal region ($N_{\textrm A}$) is calculated by taking the ratio
of the two non-tight regions ($N_{\textrm C}/N_{\textrm D}$) multiplied by the number 
of candidates in the tight, non-isolated region ($N_{\textrm B}$). 
A correction to the method is added in order to take into account the leakage of real photon events to the three background regions.
The systematic uncertainty of the method is evaluated by varying the 
criteria of tightness and isolation used to define the four regions.
This estimate also accounts for the contribution from multi-jet events, which can mimic the $\gamma+\missET$ signature if one jet is misreconstructed as a photon and one or more of the other jets are poorly reconstructed, resulting in large $\MET$.
This method is then used to evaluate the contribution of jets misidentified as photons in all analysis regions: the SRs and their associated  
four control regions, 1muCR, 2muCR, 2eleCR and PhJetCR.
The estimated contribution from this background in the SRs and the associated uncertainty are reported in Section~\ref{sec:result}.

\subsection{ Final background estimation\label{sec:finalBGestimation}}

The normalisation factors for the $W\gamma$,  $Z\gamma$ and $\gamma$+jets  backgrounds are obtained via a
profile likelihood fit (referred to as the background-only fit). 
For this fit, a likelihood function is built as the product of Poisson probability functions of the
observed and expected event yields in the control regions. 
The event yield in the corresponding SR is not considered. 
The systematic uncertainties, described in Section  \ref{sec:syst}, are treated as Gaussian-distributed nuisance parameters 
in the likelihood function. 
For each CR, the inputs to the fit are: the number of events observed in
the data, the expected numbers of $W/Z\gamma$ and $\gamma$+jets  background events, which are taken
from MC simulations and whose normalisations are free parameters in the fit, and the
number of fake-photon events obtained from the data-driven techniques. 

  Two different configurations are used for the fit: the background-only inclusive
fit, which determines the normalisations for $W\gamma$,  $Z\gamma$ and $\gamma$+jets backgrounds for each
inclusive SR independently and the background-only multiple-bin fit, which determines
the normalisations for the three exclusive SRs simultaneously. In the former case,
four CRs corresponding to a given SR are used to obtain the normalisations.
In the latter case, all ten CRs (the three leptonic CRs for each SR and the PhJetCR) associated with the three exclusive SRs are used.
  These fits are described in more detail in the following.

\subsubsection{Background-only inclusive-SR fit}
\label{inclusive_fit}

Background estimates in each inclusive SR are derived from a simultaneous fit to the respective four control regions 
(1muCR, 2muCR, 2eleCR and PhJetCR). 
The fitted values of the normalisation factors for $W/Z\gamma$ and $\gamma$+jets backgrounds (scale factors $k$) 
are reported in Table~\ref{table:met_bins_kfact}. 
  Although the PhJetCR is defined in the same way for all SRs, the $ k_{\gammajets}$ factors in the three inclusive SRs differ slightly 
  because they are  fitted together with the other CRs, which are different for the different SRs.

The inclusive-SR fit is used to set model-independent limits, as shown in Section~\ref{sec:interpretation}.

       \begin{table}[hbtp!]
       \small
       \center
               \caption{Normalisation factors (scale factors $k$) obtained from a background-only inclusive-SR fit performed in each inclusive SR (the first three columns)
               and scale factors $k'$ obtained from a background-only multiple-bin fit performed simultaneously in the three regions SRE1, SRE2 and SRI3 (the last three columns), 
               where $k'_{\gammajets}$ applies to all exclusive signal regions.
               The errors shown include both the statistical and systematic uncertainties.}
	       \label{table:met_bins_kfact}
              \begin{tabular}{ l l l l l l l l l r }
       \hline
       \hline
       Signal    &   $\met$ &~~~~$k_{W\gamma}$      &~~~~$  k_{Z\gamma}$           &    ~~~~$  k_{\gammajets}$ 	& ~~~~$  k'_{W\gamma}$& ~~~~$  k'_{Z\gamma}$  & ~~~~$  k'_{\gammajets}$  \\ 
       region    &  [GeV] & & & & & & \\\hline
SRI1                       &   $> 150$         & $1.05\pm0.09$    & $1.10 \pm  0.09 $            &  $1.07\pm0.25$              & & &    \\
SRI2                       &   $> 225$         & $1.04\pm0.11 $  & $1.14 \pm  0.13 $            &  $1.06\pm0.25 $              & & &     \\
SRI3                       &   $> 300$         & $1.04\pm0.15$   &$1.27 \pm  0.23 $             &  $1.06\pm0.24 $              &   $1.03\pm0.14$           &  $1.27\pm  0.23$         &     \\
SRE1                      &   $150$--$225$        &   &         &                                                                                        &   $1.06\pm0.10$            & $1.10\pm  0.10$          &  $1.07\pm0.25$    \\
SRE2                      &    $225$--$300$       &     &         &                                                                                       & $1.02\pm0.12$             & $1.09\pm  0.14$          &    \\

      \hline      
      \hline
       \end{tabular}
	\end{table}

\subsubsection{Background-only multiple-bin fit}
\label{multiple_fit}

A background-only multiple-bin fit is performed using simultaneously the control regions corresponding to the three signal regions SRE1, SRE2 and SRI3, 
which are mutually exclusive. The $\gamma$+jets normalisation factor is fixed in the common control region at low \met (PhJetCR), 
while the $W\gamma$ and $Z\gamma$ normalisation factors are fitted in each \met\ range separately. 
The estimated normalisation factors (scale factors $k'$) after this multiple-bin fit for each of the three SRs are also reported in Table~\ref{table:met_bins_kfact}. 
As expected, they agree within uncertainties with the scale factors $k$ obtained from the inclusive-SR fit. 

\begin{figure}[!htbp]
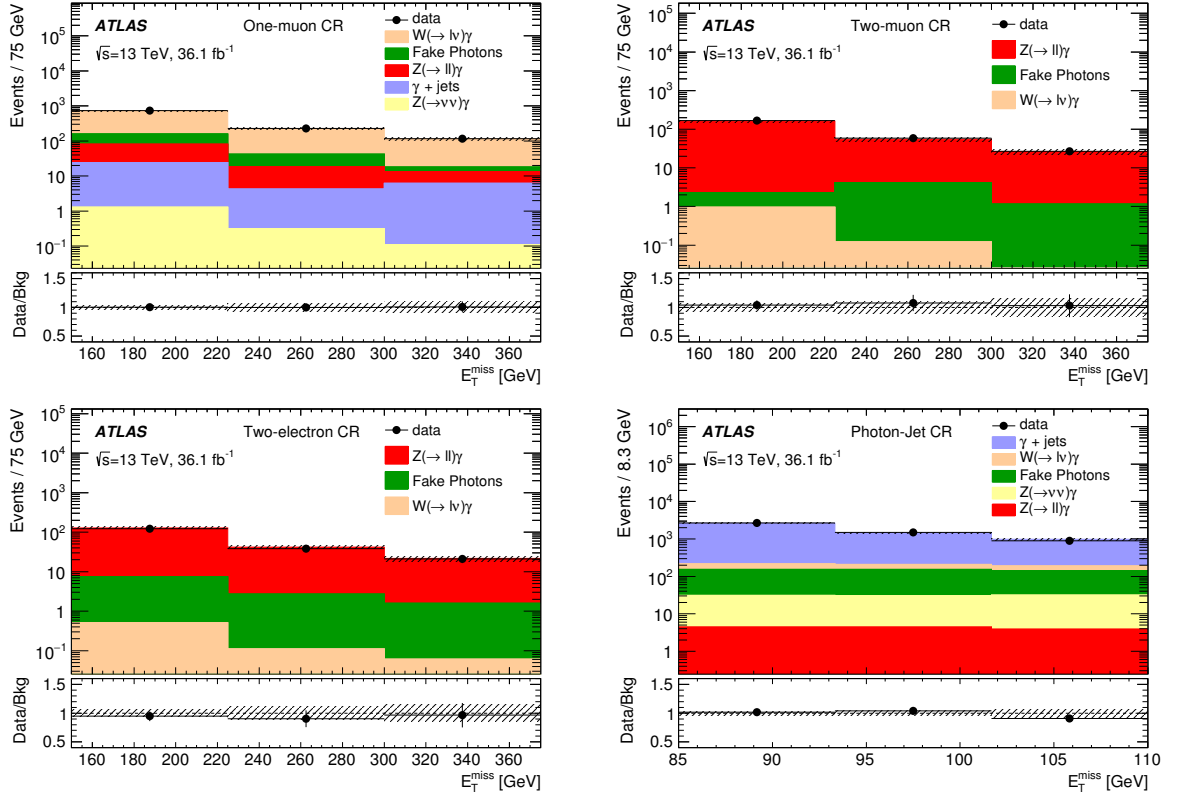
\centering
\includegraphics[width=0.49\linewidth]{{{figures/fig_02a}}}
\includegraphics[width=0.49\linewidth]{{{figures/fig_02b}}}\\
\includegraphics[width=0.49\linewidth]{{{figures/fig_02c}}}
\includegraphics[width=0.49\linewidth]{{{figures/fig_02d}}}
\caption{
Distribution of $\missET$ in data and for the expected total background in the CRs:
1muCR (top left), 2muCR (top right), 2eleCR (bottom left) and PhJetCR (bottom right).
In 1muCR and  2muCR, the muons are treated as non-interacting particles in the $\met$ reconstruction;
the electrons are handled similarly in 2eleCR.
The total background expectation is normalised using the scale factors $k'$ derived from the multiple-bin fit. 
For 1muCR, 2muCR and 2eleCR, overflows are included in the third bin. 
The error bars are statistical, and the dashed band includes statistical and systematic uncertainties determined by the multiple-bin fit. 
The lower panel shows the ratio of data to expected background event yields.}
\label{fig:CRall}
\end{figure}

Post-fit distributions of $\missET$ in the four control regions are shown in Figure~\ref{fig:CRall}.
The scale factors $k'$ from the multiple-bin fit are used for the different $\missET$ ranges to produce these figures.
These distributions illustrate the contribution from the different background processes. 

The multiple-bin fit is used to set exclusion limits  in the models studied,  if no excess is found in the data, as discussed in Section~\ref{sec:interpretation}.

%% file: Results.tex
Table \ref{tab:SR_yields} presents the observed number of events and the SM background 
predictions in SRI1 that is the most inclusive signal region with the lowest $\met$ threshold,  as obtained from the simultaneous inclusive-SR fit to its CRs. 
The corresponding number of events is shown in the three lepton CRs and in PhJetCR.
 For the SM predictions both the statistical and systematic uncertainties, described in Section~\ref{sec:syst}, are included.

Table~\ref{tab:SRall_yields} shows the observed number of events and the total SM background prediction after the fit in all signal regions.
For SRI1, SRI2 and SRI3 regions the expected SM event yields are obtained from the inclusive-SR fit to each region; 
for SRE1 and SRE2 regions the expected SM event yields are obtained from the multiple-bin fit to the regions SRE1, SRE2 and SRI3.
The expected SM event yields in SRI3 are the same when obtained from the multiple-bin fit. 
The numbers of observed events in the corresponding lepton CRs for each SR are also reported. 

\begin{figure}[!htbp]\centering
\includegraphics[width=0.49\linewidth]{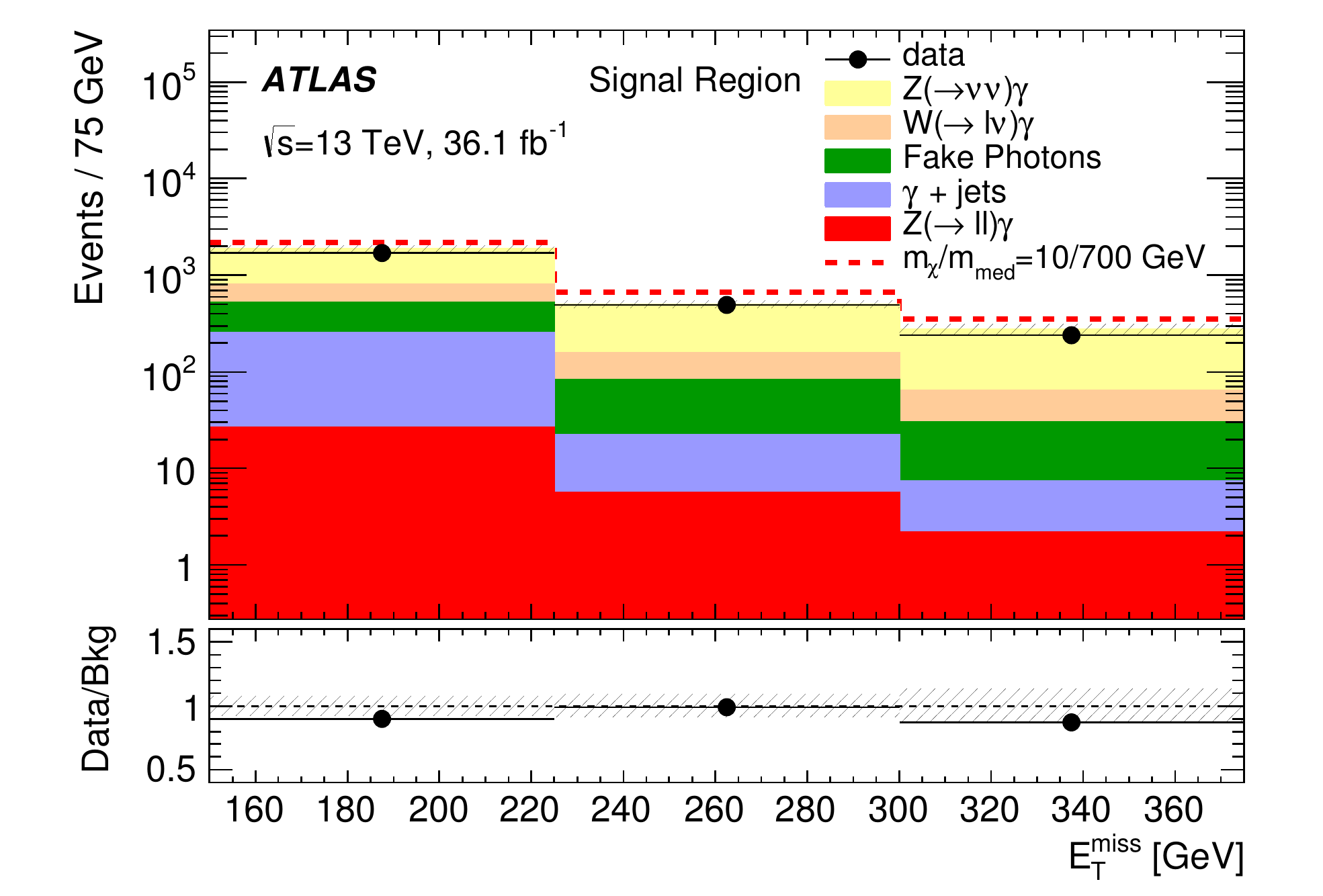}
\includegraphics[width=0.49\linewidth]{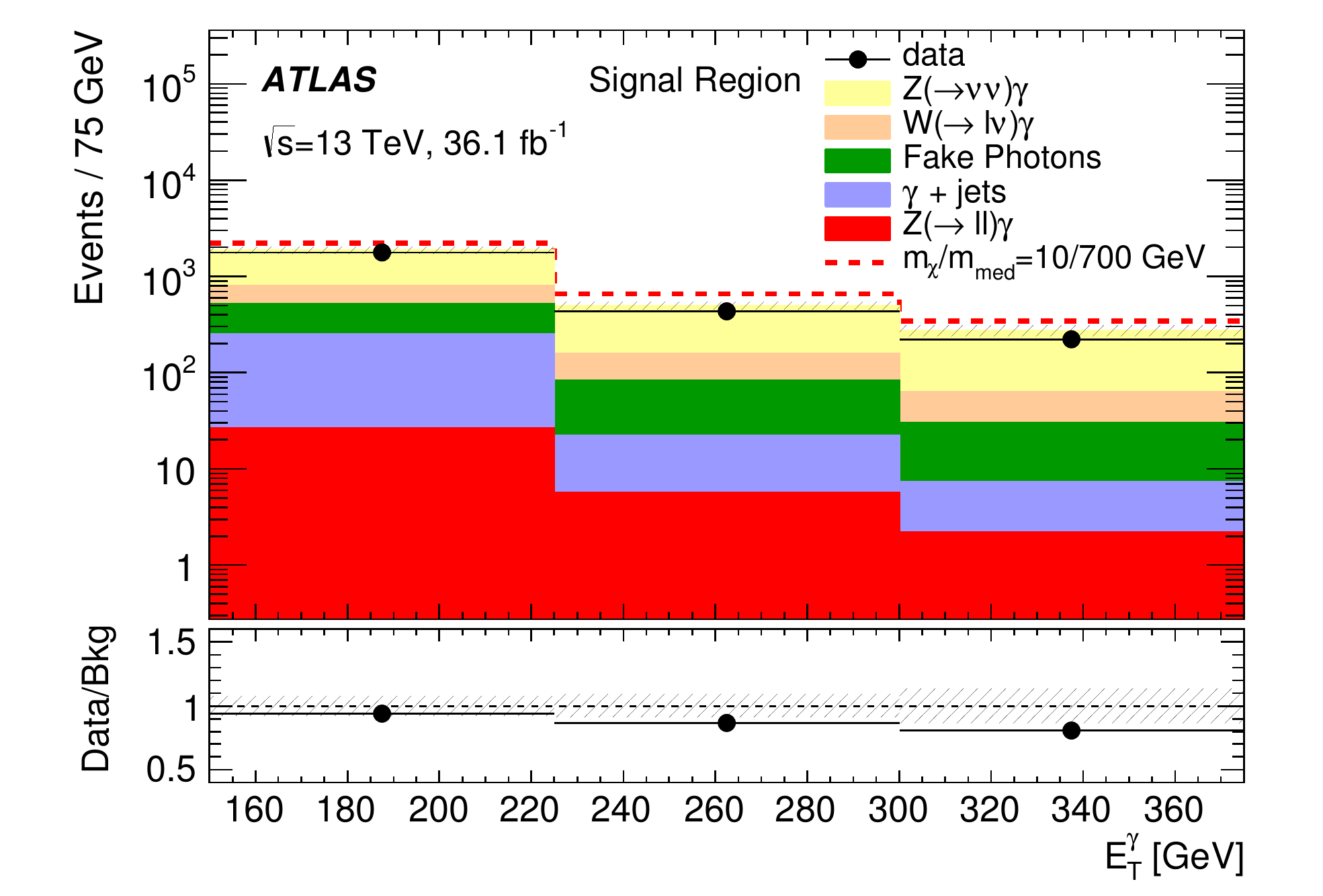}
\caption{
Distribution of $\missET$ (left) and  of  $\phPT$ (right) in the signal regions for data and for the expected total background; 
the total background expectation is normalised using the scale factors $k'$ derived from the multiple-bin fit. 
Overflows are included in the third bin. 
The error bars are statistical, and the dashed band includes statistical and 
systematic uncertainties determined by the fit. 
The expected yield of events from the simplified model with $m_{\chi} = 10\GeV$ and an axial-vector mediator of mass $m_{\textrm{med}} = 700\GeV$ with $g_{q}=0.25$ and $g_{\chi}=1.0$ is stacked on top  of the background prediction.
The lower panel shows the ratio of data to expected background event yields.
}
\label{fig:SR_event}
\end{figure}

The post-fit $\missET$ and $\phPT$ distributions are shown in Figure~\ref{fig:SR_event} after applying the 
scale factors $k'$ from the multiple-bin fit.
 Only the $\missET$ distribution is used in the multiple-bin fit, as discussed in  Section \ref{sec:interpretation}.

\begin{table}[!htbp]
\begin {center}
\caption{
Observed event yields in 36.1 \ifb~of data compared to  expected yields from SM
backgrounds in the signal region SRI1 and in its four control regions (CRs), 
as predicted from the simultaneous  fit to CRs of SRI1 (see text).  
The MC yields before the fit are also shown.
The uncertainty includes both the statistical and systematic uncertainties described in
Section~\ref{sec:syst}.
The uncertainty on the pre-fit background is the pre-fit uncertainty, while the uncertainties on the fitted background are post-fit uncertainties. The latter are constrained by the fit as the use of control regions to normalise the dominant backgrounds allows to partially cancel some systematic uncertainties 
(see Section~\ref{sec:syst} for more details).
The individual uncertainties can be correlated and do not necessarily 
add in quadrature to equal the total background uncertainty.
The total fitted background does not match exactly the sum of the individual   
contributions because of the rounding applied.
}

\label{tab:SR_yields}
\sisetup{round-mode=places, round-precision=2,
  retain-explicit-plus=true, group-digits = integer, group-minimum-digits=4}
\begin{tabular}{l D{,}{\pm}{-1} D{,}{\pm}{-1} D{,}{\pm}{4.4} D{,}{\pm}{4.4} D{,}{\pm}{4.4} }
\toprule
\toprule
& \multicolumn{1}{c}{ SRI1}& \multicolumn{1}{c}{ 1muCR}& \multicolumn{1}{c}{ 2muCR}&
\multicolumn{1}{c}{ 2eleCR}& \multicolumn{1}{c}{ PhJetCR}\\

\midrule
\multicolumn{1}{l}{ Observed events} & \multicolumn{1}{c}{2400} & \multicolumn{1}{c}{
1083} &  \multicolumn{1}{c}{ 254} &  \multicolumn{1}{c}{ 181} &  \multicolumn{1}{c}{
5064} \\
\midrule

\multicolumn{1}{l}{Fitted Background}                          &2600,160   & 1083,33     &  243,13       &193,10 &5064,80\\
\midrule
\multicolumn{1}{l}{$Z(\rightarrow \nu\nu)\gamma$ }     & 1600,110    & 1.7,0.2       &\multicolumn{1}{c}{ $-$ }     &\multicolumn{1}{c}{ $-$  }    &81,6\\
\multicolumn{1}{l}{ $W(\rightarrow \ell\nu)\gamma$}    & 390,24        & 866,40       &1.1,0.3         &0.7,0.1     &163,9\\
\multicolumn{1}{l}{$Z(\rightarrow \ell\ell)\gamma$}       & 35,3            & 77,5           &233,13         &180,10         &13,1\\                      
\multicolumn{1}{l}{$\gammajets$ }                                & 248,80        & 33,8           &\multicolumn{1}{c}{ $-$ }    &\multicolumn{1}{c}{ $-$ }    &4451,80\\
\multicolumn{1}{l}{ Fake photons from electrons}         & 199,40        & 17,3           &0.50,0.13     &0.09,0.04     &72,14\\  
\multicolumn{1}{l}{Fake photons from jets }                  & 152,22        & 88,19         &7.9,3.8         &12,5           &284,28\\   
\midrule
\multicolumn{1}{l}{Pre-fit background}                          & 2400,200     & 1025,72    &  218,15        &181,13         & 4800,1000\\

\bottomrule
\bottomrule
\end{tabular}
\end {center}
\end{table}

\begin{table}[!htbp]
\begin {center}
\caption{
Observed event yields in 36.1 \ifb~of data
compared to  expected yields from SM
backgrounds in all signal regions, as predicted from the simultaneous  fit to their respective CRs (see text).  
The first three columns report the yields obtained from the inclusive-SR fit, while the two last
columns report the yields obtained from the multiple-bin fit.
The uncertainty includes both the statistical and systematic uncertainties described in
Section~\ref{sec:syst}.
The uncertainties are post-fit uncertainties and are constrained by the fit as the use of control regions to normalise the dominant backgrounds allows to partially cancel some systematic uncertainties (see Section~\ref{sec:syst} for more details).
The individual uncertainties can be correlated and do not necessarily 
add in quadrature to equal the total background uncertainty.
The observed number of events in the four CRs relative to each SR is also shown.
The total fitted background does not match exactly the sum of the individual   
contributions because of the rounding applied.
}

\label{tab:SRall_yields}
\begin{tabular}{l D{,}{\pm}{-1} D{,}{\pm}{-1} D{,}{\pm}{4.4} D{,}{\pm}{4.4}
D{,}{\pm}{4.4} }
\toprule
\toprule
& \multicolumn{1}{c}{ SRI1}& \multicolumn{1}{c}{ SRI2}& \multicolumn{1}{c}{ SRI3}&
\multicolumn{1}{c}{ SRE1}& \multicolumn{1}{c}{ SRE2}\\
\midrule
\multicolumn{1}{l}{Observed events} & \multicolumn{1}{c}{2400} & \multicolumn{1}{c}{729} &  \multicolumn{1}{c}{236} &  \multicolumn{1}{c}{1671} &  
\multicolumn{1}{c}{493} \\
\midrule
\multicolumn{1}{l}{Fitted Background}                               &  2600,160  & 765,59  & 273,37    &   1900,140     & 501,44\\
\midrule
\multicolumn{1}{l}{$Z(\rightarrow \nu\nu)\gamma$ }          & 1600,110   & 543,54   & 210,35    &   1078,89     & 342,41\\
\multicolumn{1}{l}{ $W(\rightarrow \ell\nu)\gamma$}         & 390,24       & 109,9     & 33,4         &   282,22     & 75,8\\
\multicolumn{1}{l}{$Z(\rightarrow \ell\ell)\gamma$}            & 35,3           & 7.8,0.8   &2.2,0.4    &   27,3     & 5.7,0.7\\                      
\multicolumn{1}{l}{$\gammajets$ }                                     & 248,80       & 22,7       & 5.2,1.0   &   225,80     & 17,6 \\
\multicolumn{1}{l}{ Fake photons from electrons}              & 199,40       & 47,11     & 13,3         &    152,28    & 34,8\\ 
\multicolumn{1}{l}{Fake photons from jets }                       & 152,22       & 37,15     &   \multicolumn{1}{c}{$9.7^{+10}_{-9.7}$}  &   115,24     & 27,9\\   
\midrule
\multicolumn{1}{l}{Observed events in 1muCR} &\multicolumn{1}{c}{1083} &\multicolumn{1}{c}{343}  &\multicolumn{1}{c}{116}  &\multicolumn{1}{c}{740}  &\multicolumn{1}{c}{227}\\
\midrule
\multicolumn{1}{l}{Observed events in 2muCR} &\multicolumn{1}{c}{254}  &\multicolumn{1}{c}{86}    &\multicolumn{1}{c}{27}    &\multicolumn{1}{c}{168}  &\multicolumn{1}{c}{59}\\
\midrule
\multicolumn{1}{l}{Observed events in 2eleCR}  &\multicolumn{1}{c}{181}  &\multicolumn{1}{c}{59}   & \multicolumn{1}{c}{21}    & \multicolumn{1}{c}{122}  &\multicolumn{1}{c}{38}\\
\midrule
\multicolumn{1}{l}{Observed events in PhJetCR}  &\multicolumn{1}{c}{5064}  &\multicolumn{1}{c}{5064}   & \multicolumn{1}{c}{5064}    & \multicolumn{1}{c}{5064}  &\multicolumn{1}{c}{5064}\\
\bottomrule
\bottomrule
\end{tabular}
\end {center}
\end{table}

%% file: Systematics.tex
The systematic uncertainties are treated as Gaussian-distributed nuisance
 parameters in the likelihood function fitted to obtain the final
 background predictions in the SRs, as described in Section~\ref{sec:BGEstimation}.
The fit  provides constraints on many sources of systematic uncertainty, as the normalisations of the dominant background 
processes are fitted parameters; only the uncertainties affecting the extrapolation between CRs and SRs therefore remain important.

The fitted uncertainties are presented as percentages of the total background predictions in SRs.
The total background prediction uncertainty, including systematic and statistical contributions, 
varies from $6.1\%$ to $14\%$  for the five SRs, dominated by the  statistical uncertainty in the control regions, 
which varies from  approximately $4.3\%$ to $10\%$.

The relevant uncertainties (giving a contribution of more than 0.1\% in at least one SR) 
are summarised in Table~\ref{tab:syst} for all SRs. 

Aside from the uncertainty due to the statistical precision from the CRs, 
the largest relative systematic uncertainties are due to the uncertainty in the rate of fake photons from jets, which contributes $1.3\%$--$5.3\%$,
increasing for SRI3 and SRE2 because of the smaller number of events available for the estimation, 
and to the uncertainty in the jet energy scale, which contributes $1.4\%$--$5.6\%$, decreasing in the regions with larger $\MET$. 
The systematic uncertainty due to the modelling of the $\gamma$+jets background, which affects the extrapolation of this background from 
the PhJetCR to the SRs, is evaluated by  independently varying the following four parameters with respect to the nominal values used in the MC generation:
the renormalisation, factorisation and resummation scales by factors of 2.0 and 0.5, 
and the CKKW matching scale~\cite{CKKW} to 15 GeV and 30 \GeV\ (the nominal value being 20 \GeV). 
 For the $W/Z\gamma$ backgrounds, the lepton identification/reconstruction efficiency uncertainties are propagated from the leptonic CRs to the SRs in terms of veto efficiency uncertainties.
After the fit,  the uncertainty in the luminosity \cite{lumi2013} is found to have a negligible impact on the background estimation.
The uncertainties due to the PDF have an impact on  the $W/Z\gamma$ samples in each region but the effect on normalisation is largely 
absorbed in the fit, so their impact is negligible. 

For the signal-related systematic uncertainties, the uncertainties  due to PDF are evaluated 
following the PDF4LHC recommendations~\cite{PDF4LHC} and using a reweighting procedure implemented in the LHAPDF Tool~\cite{LHAPDF}, 
while uncertainties due to the scales are evaluated by varying the renormalisation and 
factorisation  scales by factors of 2.0 and 0.5 with respect to the 
nominal values used in the MC generation.
The uncertainties in initial- and final-state radiation, due to the 
choice of parton shower and multiple-parton-interaction parameters used with $\pythiaeight$ are 
estimated by generating MC samples with the alternative
tunes described in Ref.~\cite{ATL-PHYS-PUB-2014-021}. 
The PDF, scale and tune each induce uncertainties of up to about 5\% in
the acceptance (and cross section) in the simplified DM models.

\begin{table}[!hbtp]
\centering
\small
\caption{
Breakdown of the relevant uncertainties in the background estimates for all SRs.
The uncertainties are given relative to the expected total background yield after
 the inclusive-SR fit for SRI1, SRI2 and SRI3 and after the multiple-bin fit for SRE1 and SRE2.
 The total statistical uncertainty is dominated by the  statistical uncertainty in CRs. 
The individual uncertainties can be correlated and do not necessarily 
add in quadrature to equal the total background uncertainty.
 }
 
\label{tab:syst}
\begin{tabular}{lrlrlrlrlr}
\hline\hline
                     & SRI1& SRI2 & SRI3 & SRE1& SRE2   \\
 \hline
Total background & 2600 & 765&273& 1900& 501\\ 
Total (statistical+systematic) uncertainty  & $6.1\%$ &7.7\%&14\% &7.7\% & 8.8\% \\ 
Statistical uncertainty only & $4.3\%$ &6.2\%&10\% &5.5\% & 7.8\%\\ 

\hline
Jet fake rate  (Section~\ref{sec:bJetfake})                                                                       & $1.3\%$         &$ 3.0\%$        &$ 5.3\%$       &   1.7\%    & 3.3\%  \\    
Electron fake rate   (Section~\ref{sec:bElefake})                                                              & $ 1.5\%$         &$ 1.5\%$        &$ 1.2\%$       &   1.5\%    & 1.6\%  \\             
Jet energy scale~\cite{PERF-2016-04}                                               & $4.1\%$         & $1.9\%$         & $1.4\%$       &  5.6\%     & 0.6\%\\
Jet energy resolution~\cite{JER3}                                          & $0.7\%$         &  $0.2\%$     & $0.5\%$       &  0.8\%        & 0.3\% \\
\met\ soft-term scale and resolution~\cite{MET_Paper_8TeV}                                  & $0.9\%$        &$0.4\%$         &$0.7\%$        &   1.1\%       & 1.0\% \\   
Muon reconstruction/isolation efficiency~\cite{Muons13TeV}                              & $1.4\%$         &$ 1.3\%$        &$ 1.6\%$       &   1.3\%      & 1.4\%    \\  
Electron reco/identification/isolation efficiency~\cite{PERF-2013-03}     &  $1.0\%$         & $1.3\%$         & $1.3\%$       & 0.8\%      & 1.2\%       \\
Electron and photon energy scale~\cite{PERF-2013-05}                                       & $0.2\% $         &$0.5\%$         &$0.4\%$      &  $<0.1\% $   &  $<0.1\% $ \\
Electron and photon energy resolution~\cite{PERF-2013-05}                              & $<0.1\% $      & $0.3\%$         & $0.2\%$       &  0.1\%     & 1.0\%   \\        
Photon efficiency~\cite{PhId}                                                                  & $0.1\%$       & $1.0\%$      & $<0.1\%$     &   0.2\%     & $<0.1\%$\\
$\gamma+$jets modelling                                                                    & $1.5\%$       & $0.3\%$      & $0.3\%$     &   2.3\%     & 0.4\%  \\
$\langle \mu \rangle$ distribution in MC simulation  (Section~\ref{sec:simsample})                                & $1.3\%$         & $0.3\%$        & $0.9\%$       &   1.7\%     & 0.3\%\\
\hline
\hline
\end{tabular}
\end{table}

%% file: Interpretation.tex
The event yields observed in data are consistent within uncertainties with the predicted SM background event yields  in all inclusive SRs, 
as shown in Table~\ref{tab:SRall_yields}.
The results from the SRs shown in Section~\ref{sec:result} are therefore interpreted in terms of exclusion limits in models that would produce an excess of  $\gamma+\missET$ events.
 Upper bounds are calculated using a one-sided profile likelihood ratio and the $CL_S$ 
technique~\cite{Read:2002hq,Junk:1999kv}, evaluated using the asymptotic approximation~\cite{Cowan:2010js}.
The likelihood fit includes both the SRs and their CRs.

The upper limits on the visible cross section, 
defined as the product of the cross section times the acceptance times the reconstruction efficiency defined in a fiducial region, $\sigma\times A \times \epsilon$, 
of a potential  BSM signal, are obtained from the three inclusive SRs.
The value of $A$ for a particular model is computed by applying the same selection criteria as in the SR but at the 
particle level; in this computation \metv\ is given by the vector sum of the transverse momenta of all non-interacting particles. 
The $A$ values  with the selection for SRI1 or SRI2 or SRI3  are reported in Table~\ref{tab:xsec_vis_fid} for the simplified DM models; 
the lowest values are found in models with low-mass off-shell mediators and the highest values in models with 
high-mass on-shell mediators.
 The observed and expected upper limits, at  95\% confidence level (CL),
  on the fiducial cross section, defined as $\sigma\times A$ are shown in Table~\ref{tab:xsec_vis_fid}.
 They are computed by dividing the limit on the visible cross section  by the fiducial 
reconstruction efficiency $\epsilon$  shown in the same table; 
as in the case of the acceptance, the efficiency is smaller in the high-$\met$ bins.
The lowest efficiency for each signal region is used in a conservative
way to set the fiducial cross-section limit.
These limits can be extrapolated to models producing $\gamma+\missET$ 
 events once $A$ is known, assuming that the conservative efficiency  applies. 

\begin{table}[!htbp]
\small
\begin {center}
\caption{The observed and expected limits at 95\% CL 
on the fiducial cross section $\sigma\times A$.
The values of the fiducial reconstruction efficiency ($\epsilon$), which is used for the calculation of the fiducial cross section, and of the acceptance ($A$) are also shown.
}

\label{tab:xsec_vis_fid}
\begin{tabular}{c D{,}{\pm}{-1} D{,}{\pm}{-1} D{,}{\pm}{4.4} }
\toprule
\toprule
\multicolumn{1}{l} {} &   \multicolumn{3}{c} {$\sigma\times A$ limit [fb]} \\
\midrule
\multicolumn{1}{l}{Region} & \multicolumn{1}{c}{ SRI1}& \multicolumn{1}{c}{ SRI2}& \multicolumn{1}{c}{ SRI3}\\
\midrule
\multicolumn{1}{l}{95\% CL observed} & \multicolumn{1}{c}{ 7.0}& \multicolumn{1}{c}{3.7 }& \multicolumn{1}{c}{ 2.3}\\
\multicolumn{1}{l}{95\% CL expected} & \multicolumn{1}{c}{10.6 }& \multicolumn{1}{c}{4.5 }& \multicolumn{1}{c}{3.0 }\\
\multicolumn{1}{l}{95\% CL expected ($\pm1\sigma$)} & \multicolumn{1}{c}{ 14.5, 7.7}& \multicolumn{1}{c}{6.2, 3.3 }& \multicolumn{1}{c}{4.2, 2.2 }\\
\multicolumn{1}{l}{$A$ [\%]} &  \multicolumn{1}{c}{ 14--48}& \multicolumn{1}{c}{5--31 }& \multicolumn{1}{c}{ 2--19}\\
\multicolumn{1}{l}{$\epsilon$ [\%]} &  \multicolumn{1}{c}{ 84--95}& \multicolumn{1}{c}{ 73--86}& \multicolumn{1}{c}{ 64--85}\\
\bottomrule
\bottomrule
\end{tabular}
\end {center}
\end{table}

The expected limit on the signal strength in the simplified DM model is computed 
 with the inclusive-SR fit for the various inclusive regions 
and with the multiple-bin fit in order to determine which strategy to adopt for limit setting. 
While SRI1 is the inclusive SR that gives  the most stringent expected limits at very low DM/mediator masses, 
SRI2 is the inclusive SR providing the most stringent  limits in the rest of the parameter space; 
SRI3, which has larger uncertainties, is not able to set better expected constraints on high-mass models in spite of their harder \met\ spectra. 
The multiple-bin fit, making use of the expected signal distribution in $\met$ by combining the information from the various exclusive SRs, 
allows more stringent expected limits to be set than in any of the inclusive signal regions.

The results are presented  for both the axial-vector and vector mediators  using different couplings
 to show the complementarity to the direct searches in $X+\missET$ events and  the searches looking for the mediator, 
 such as dijet or dilepton resonance searches, as recommended in Ref.~\cite{DMwg}.
 Four models are considered with different mediators and different couplings to quarks, to DM particles, and to leptons, and these models are
summarised in Table~\ref{models}.
As the choices of mediators and of couplings  only affect the signal cross section and not the acceptance for signal events, 
the events generated for the axial-vector mediator with $g_q$ = 0.25, $g_{\chi}$ =1 and $g_{\ell} = 0$ (model A1), 
described in Section~\ref{sec:simsample}, can be re-scaled in order to obtain results for the other three models.

When placing limits in specific models, the signal-related systematic 
uncertainties estimated as described in Section \ref{sec:syst}
affecting $A \times \epsilon$ (PDF, scales, initial- and 
final-state radiation) are included in the statistical analysis,  while 
the uncertainties affecting the cross section (PDF, scales) are 
indicated as bands around the observed limits and written as 
$\sigma_{\textrm{theo}}$. 

\begin{figure}[!htbp]
    \centering
	\includegraphics[width=0.49\textwidth]{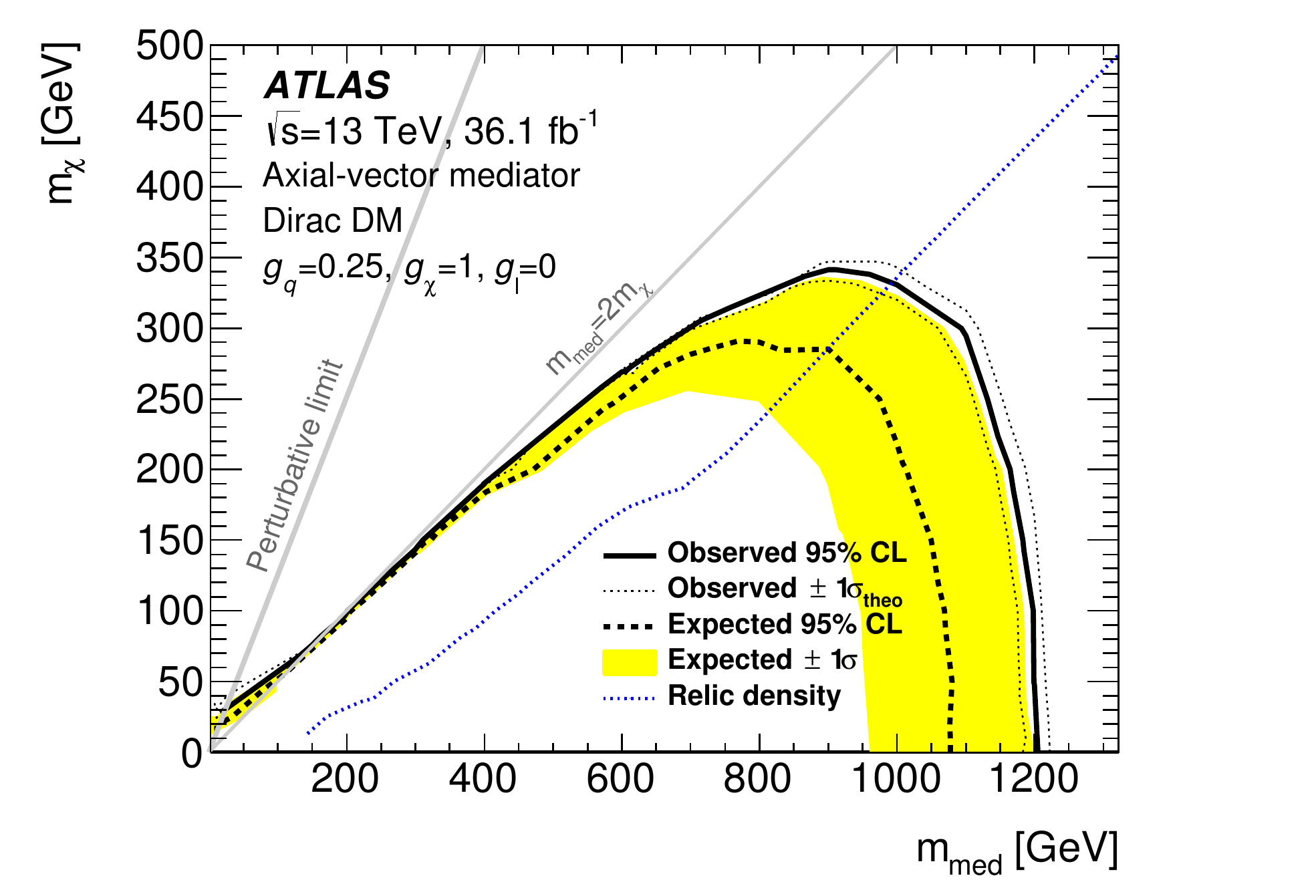}
	\includegraphics[width=0.49\textwidth]{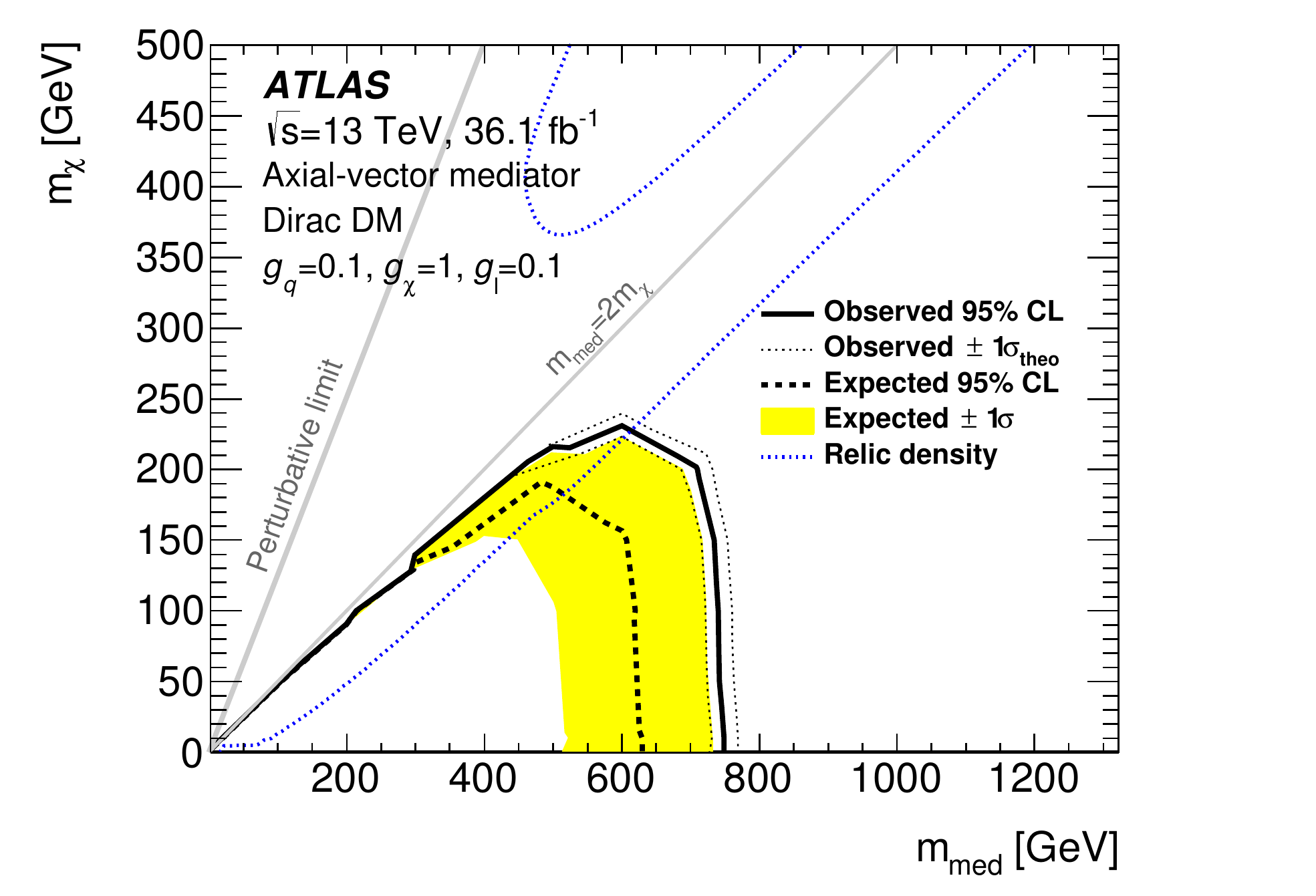}	\\
	\includegraphics[width=0.49\textwidth]{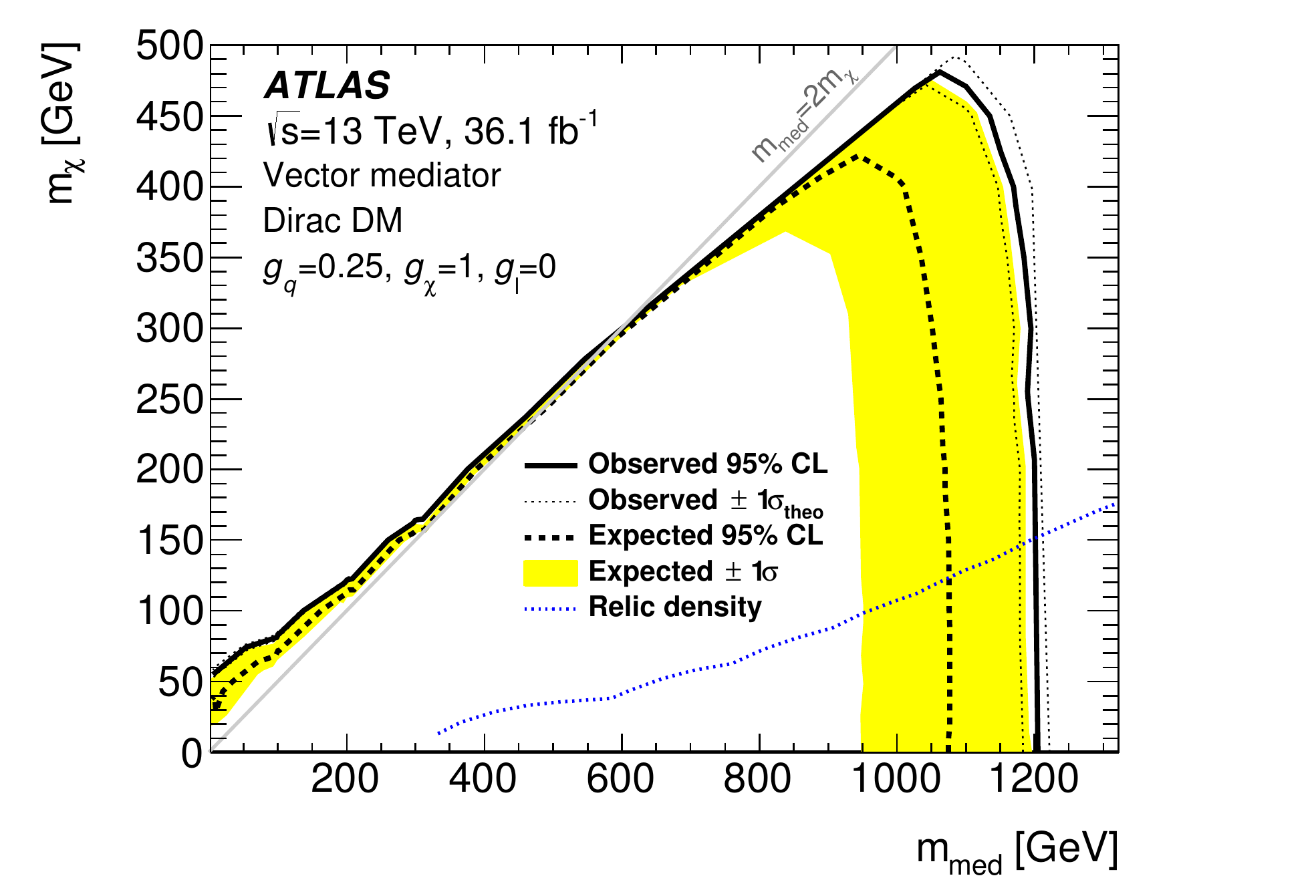}	
        \includegraphics[width=0.49\textwidth]{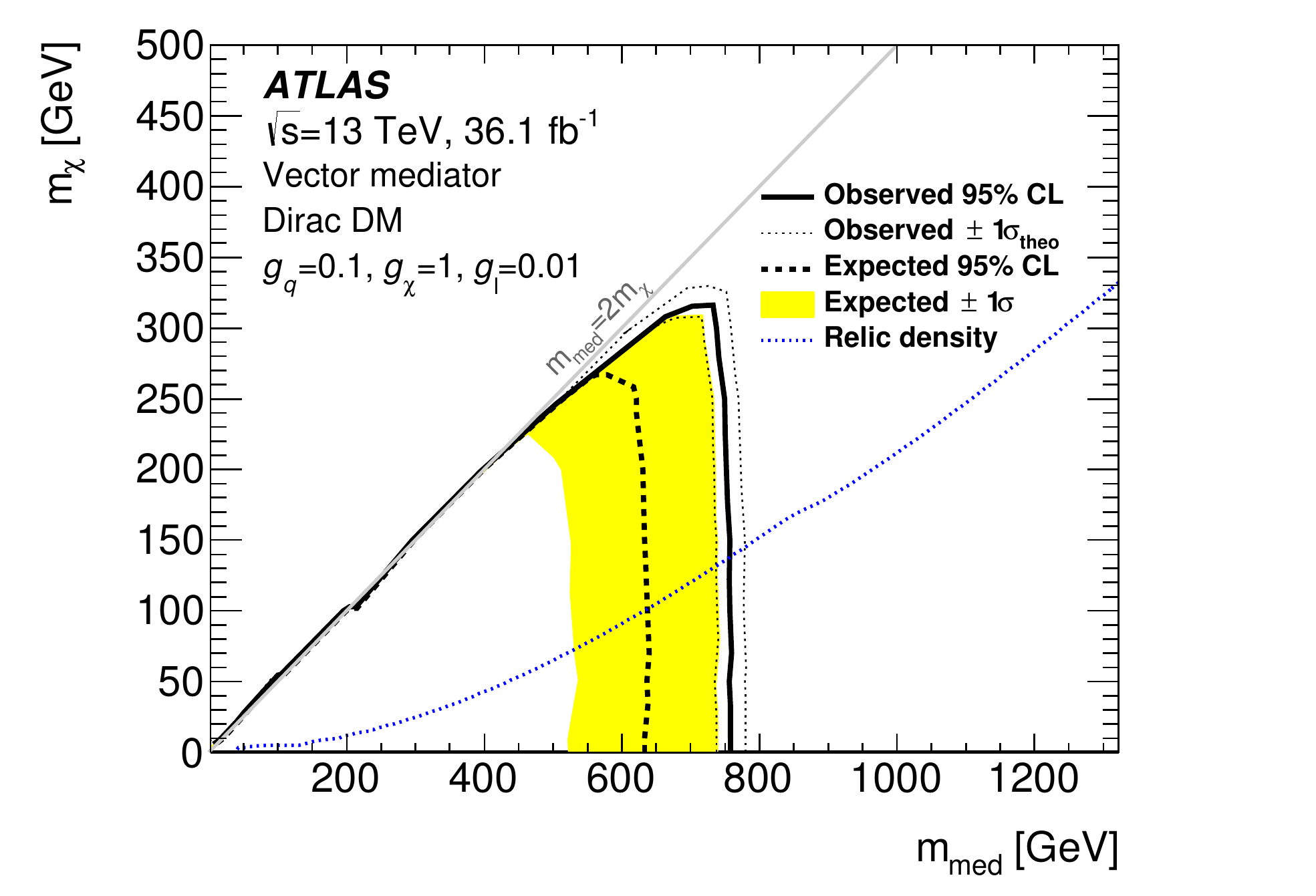}	\\	
	\caption{The observed and expected 95\% CL exclusion contours for a simplified model of dark-matter production  
	involving an axial-vector operator, Dirac DM and couplings $g_q$ = 0.25, $g_{\chi}$ = 1  and $g_\ell$ = 0 
	 as a function of  the dark-matter mass $m_{\chi}$ and the  mediator mass $m_{\textrm{med}}$ (upper left). 
	 The plane under the limit curve is excluded.
	  The same is shown for an axial-vector operator with couplings $g_q$ = 0.1, $g_{\chi}$ = 1 and  $g_\ell$ = 0.1 (top right), for a vector operator with couplings $g_q$ = 0.25, $g_{\chi}$ =1 and $g_\ell$ = 0  (bottom left) and 
	  for a vector operator with couplings $g_q$ = 0.1, $g_{\chi}$ = 1 and $g_\ell$ = 0.01 (bottom right). 
	 The region on the left is excluded by the perturbative limit which is relevant for axial-vector mediators~\cite{dmrecomm}.
	  The relic density curve \cite{Ade:2015xua, DMwg} is also shown: at higher mediator masses, the DM would be overabundant; at lower values, it would be underabundant; for the axial-vector
scenario shown in the upper right figure, the region above the relic
density curve at high dark-matter masses is also overabundant.
	  }
	\label{fig:dmA_exclusion}
\end{figure}

Simplified models with explicit mediators are valid for all values of  momentum transfer in $pp$ collisions~\cite{dmforum}. 
 Figure~\ref{fig:dmA_exclusion} (top left) shows the observed and expected contours 
corresponding to a 95\% CL exclusion as a function of $m_{\textrm{med}}$ and  $m_{\chi}$ 
for the simplified model A1.
 The region of the plane under the limit curves is excluded.
 The region not allowed due to perturbative unitarity violation is to the left of the line defined by 
 $m_{\chi} = \sqrt{\pi/2}m_{\textrm{med}}$~\cite{Perturb}.
 The line corresponding to the DM thermal relic abundance measured by
 the Planck collaboration \cite{Ade:2015xua} is also indicated in the
 figure; it is obtained as detailed in Ref.~\cite{DMwg}.
Figure~\ref{fig:dmA_exclusion} (top right) shows the contours for the A2 model, 
while Figure~\ref{fig:dmA_exclusion} (bottom left) and (bottom right) show the contours for the V1 and V2 models, respectively.
The search excludes mediator masses below the values reported in Table~\ref{models} for $\chi$ masses below
the values reported in the same table.
The limits for the model A1 are more stringent than the limits obtained with the 2015 data only~\cite{Monophoton2015}, which  
excluded mediator masses below $710\GeV$ for $\chi$ masses below $150\GeV$.

         \begin{table}[hbtp!]
       \center
               \caption{Observed limits at $95\%$ CL on the mediator mass and the DM particle mass for the four models considered.
               The mediators and couplings to quarks, to dark-matter particles and to leptons are specified for each model. }
	       \label{models}
              \begin{tabular}{ l l l l l l l l r }
       \hline
       \hline
      Model & Mediator    &    ~~~~$g_{q}$    &  ~~~~$g_{\chi} $    &     ~~~~$g_{\ell}$	 & Limit on $m_{\textrm{med}}$ [\GeV] & Limit on $m_{\chi}$ [\GeV]  \\ 
     & & & & & for low  $m_{\chi}$ & reaching as high as       \\ 
      \hline 
      A1       & axial-vector          &       ~~0.25       &     ~~1                &   ~~0      & ~~~1200 & ~~~340 \\
      A2       &  axial-vector         &          ~~0.1         &      ~~1          &   ~~0.1   &  ~~~~~750 & ~~~230 \\
      V1       &   vector     &        ~~0.25      &      ~~1                &    ~~0     & ~~~1200 &    ~~~480         \\
      V2       &    vector     &           ~~0.1    &      ~~1               &    ~~0.01  & ~~~~~750 &   ~~~320          \\                                                                                  

      \hline      
      \hline
       \end{tabular}
	\end{table}

\begin{figure}
    \centering
	\includegraphics[width=0.49\textwidth]{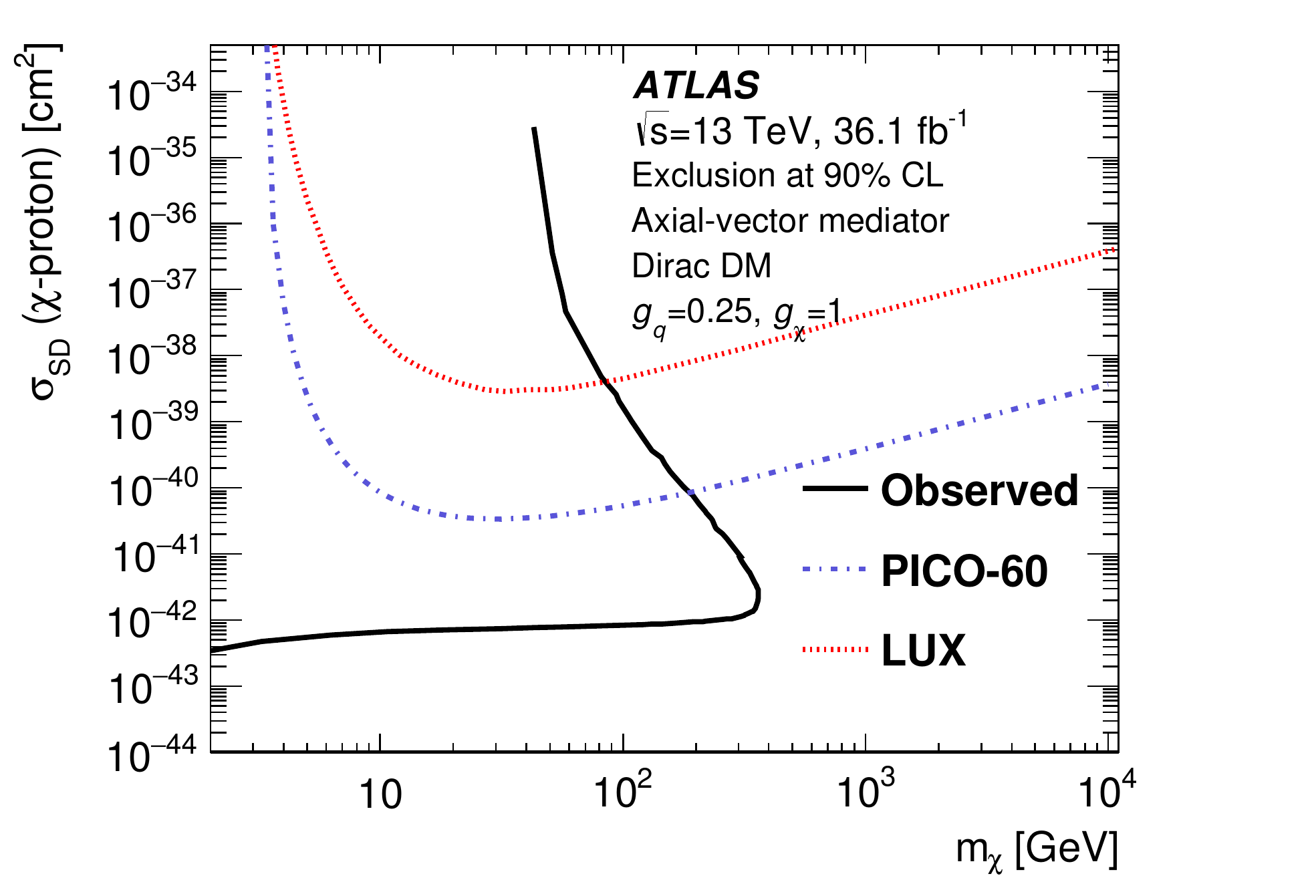}
	\includegraphics[width=0.49\textwidth]{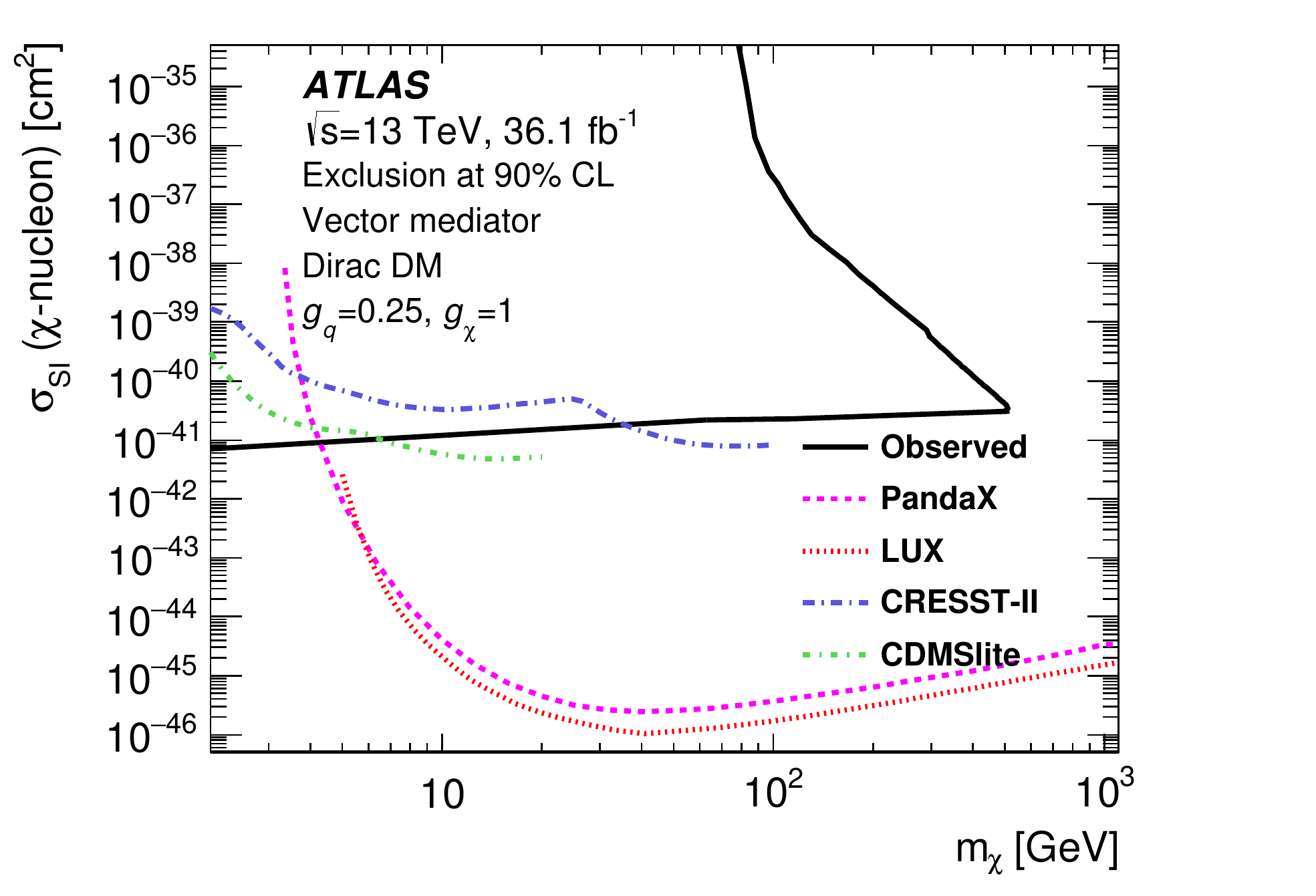}	\\

	\caption{The 90\% CL exclusion limit on the $\chi$--proton scattering cross section in a simplifed model of dark-matter production 
	involving an axial-vector operator, Dirac DM and couplings $g_q$ = 0.25, $g_{\chi}$ = 1 and $g_\ell$ = 0
        as a function of  the dark-matter mass $m_{\chi}$.
         Also shown are results at 90\% CL from two direct dark-matter search experiments~\cite{PICO_2017, LUX_new}
          (left).
          The 90\% CL exclusion limit on the $\chi$--nucleon scattering cross section in a simplifed model of dark-matter production 
	involving a vector operator, Dirac DM and couplings $g_q$ = 0.25, $g_{\chi}$ = 1 and $g_\ell$ = 0
        as a function of  the dark-matter mass $m_{\chi}$ (right); 
        also shown are results  at 90\% CL  from four direct dark-matter search experiments~\cite{PANDAX,  LUX_new2, CDMS, CRESST}.}
	\label{fig:cont_dmA}
\end{figure}

 Figure~\ref{fig:cont_dmA} (left) shows the contours corresponding to a 90\% CL exclusion translated into the plane of  $\chi$--proton spin-dependent (SD) scattering cross sections vs. $m_\chi$ 
 for the axial-vector mediator model A1.
 Bounds on the $\chi$--proton cross section are obtained following the 
procedure described in Ref.~\cite{dmrecomm}, assuming that the axial-vector mediator with couplings as in A1 
 is solely responsible for both collider $\chi$ pair production and for $\chi$--proton scattering.
  In this plane, a comparison with the result from direct DM searches~\cite{PICO_2017, LUX_new} is possible. 
The limit placed in this search extends to arbitrarily low values of $m_{\chi}$, as the acceptance at lower mass values 
is the same as the one at the lowest $m_{\chi}$ value shown here.
 The search provides stringent limits on the scattering cross section of the order of $10^{-42} {\textrm{cm}^2}$ up to $m_{\chi}$ masses of about  $300\GeV$. 
 These results allow complementary limits to be set on the $\chi$--proton scattering cross section in the low DM mass region where the  direct DM search experiments 
 have less sensitivity due to the very low energy recoils that such low-mass dark-matter particles would induce.
 Figure~\ref{fig:cont_dmA} (right) shows the limit contours in the plane of the $\chi$--nucleon spin-independent (SI) scattering cross section vs.\ $m_{\chi}$ for the vector mediator model V1 compared 
  with results of direct DM searches~\cite{PANDAX,  LUX_new2, CDMS, CRESST}. 
 In this case the limit on the scattering cross section is of the order of $10^{-41} {\textrm{cm}^2}$ up to $m_{\chi}$ masses of about  $500\GeV$.

\begin{figure}[!htbp]\centering
 	\includegraphics[width=0.54\textwidth]{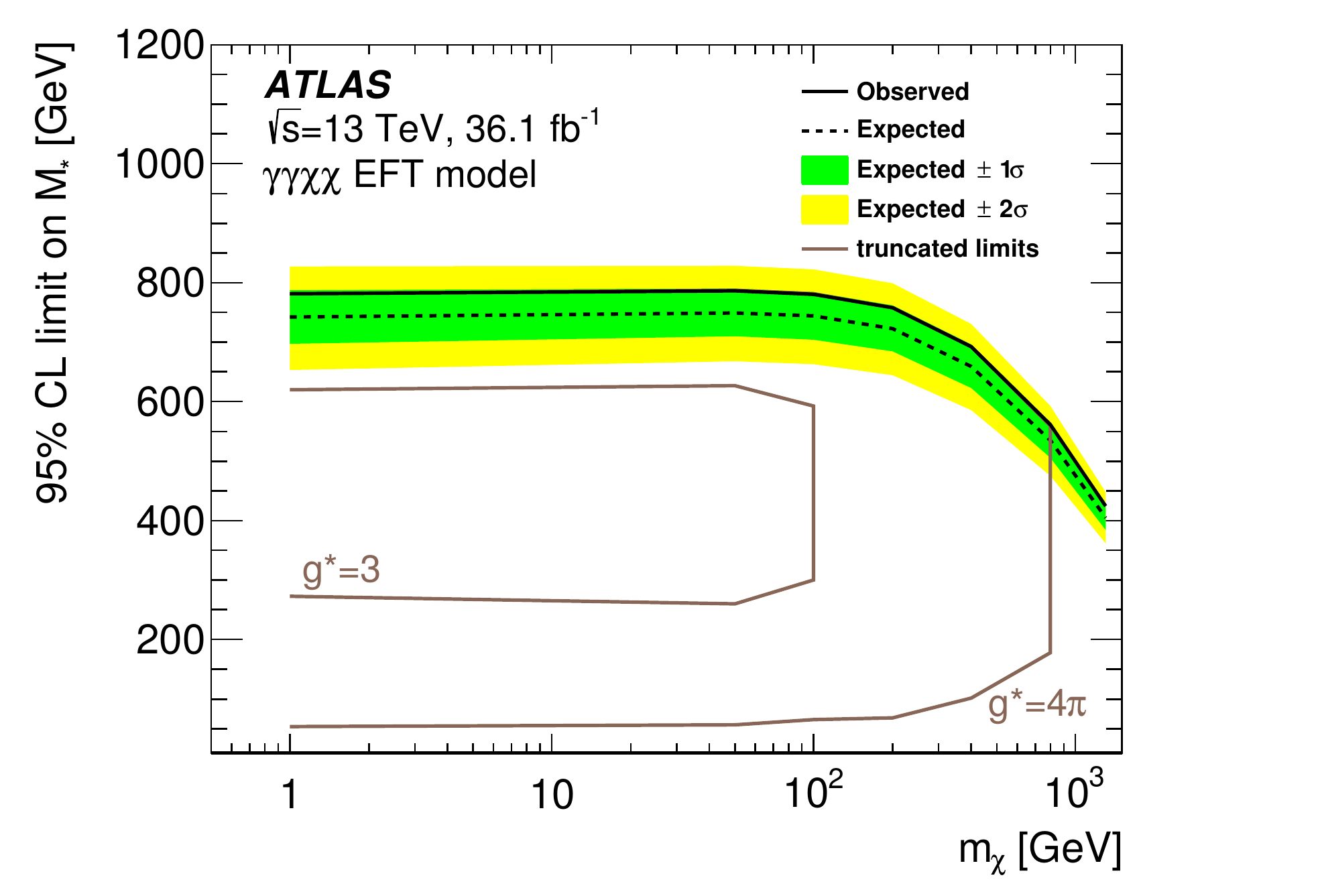}
	\caption{
	The observed and expected 95\% CL limits on $M_{*}$ for a  dimension-7 operator EFT model with a contact interaction of type $\gamma\gamma\chi\chi$  as a function of dark-matter mass $m_{\chi}$.
Results  where EFT truncation is applied are also shown, assuming representative coupling values, $g^*$, of 3  and 4$\pi$: for the maximal coupling value of 4$\pi$, the truncation
 has almost no effect; for lower coupling values, the exclusion limits are confined to a smaller area of the parameter space.
	 }
   \label{fig:Plot_schannel_Limit}
\end{figure}

In the case of the model of $\gamma\gamma\chi\bar{\chi}$ interactions, lower limits are placed on the effective 
mass scale $\Mstar$ as a function of $m_{\chi}$, as shown in Figure~\ref{fig:Plot_schannel_Limit}. 
In this model, which presents a hard $\met$ spectrum, the signal events mainly contribute to the $\met>300\GeV$ bin.
The search excludes model values of $M_*$ up to about $790\GeV$, which is a more stringent limit than the one placed 
in earlier searches~\cite{Monophoton2015}.
The EFT description is not always valid at these scales.  The effect of the truncation for two representative values of the EFT coupling, $g^{*}$, is shown in the same figure, 
assuming that the scale at which the EFT description becomes invalid ($M_{\text{cut}}$) is related to $M_{*}$ through $M_{\text{cut}}=g^{*}M_{*}$.
For the maximal coupling value of 4$\pi$, the truncation
 has almost no effect; for lower coupling values, the exclusion limits are confined to a smaller area of the parameter space.

The results are also interpreted in terms of a limit on the cross section for the production of a narrow heavy scalar $Z\gamma$ resonance produced through gluon--gluon fusion. 
Figure~\ref{fig:Zgamma} shows the observed and expected limit at 95\% CL on the production cross section of a $Z\gamma$ resonance 
as a function of  its mass. 
The limit is produced in exactly the same way as the other signal samples, where an excess of events is sought in the three exclusive signal regions by using the multiple-bin fit. 
The heavy resonances are expected to populate mainly the $\met>300$ \GeV\ signal region as they would have a hard \met\ spectrum.
The upper bound on   $m_{\ell \ell \gamma}$  applied in 2eleCR and 2muCR  (see Section~\ref{sec:bVgamma}) suppresses  the contamination 
from potential high-mass $Z\gamma$ resonances in these control regions.
Limits on such a resonance were also placed by bump searches in the very sensitive dileptonic channel and the hadronic channel for masses below and above 1.5 \TeV, respectively~\cite{Zgamma}. 
Although the $Z$ boson branching ratio to neutrinos is higher than to charged leptons, the presence of $\met$ makes the search in this channel much less sensitive than in the dileptonic channel;  
the region of interest for the analysis discussed here lies at higher masses, where it can complement the searches using $Z$ boson hadronic decays whose limits, obtained with 3.2 \ifb, are reported in the same figure.
The observed (expected) limits at $95\%$ CL on the production of a  $Z\gamma$  resonance are 
$26$ fb and $43$  fb ($32$ fb and $58$ fb) for masses of $2$ and $5\TeV$, respectively.

\begin{figure}[!htbp]\centering
 	\includegraphics[width=0.54\textwidth]{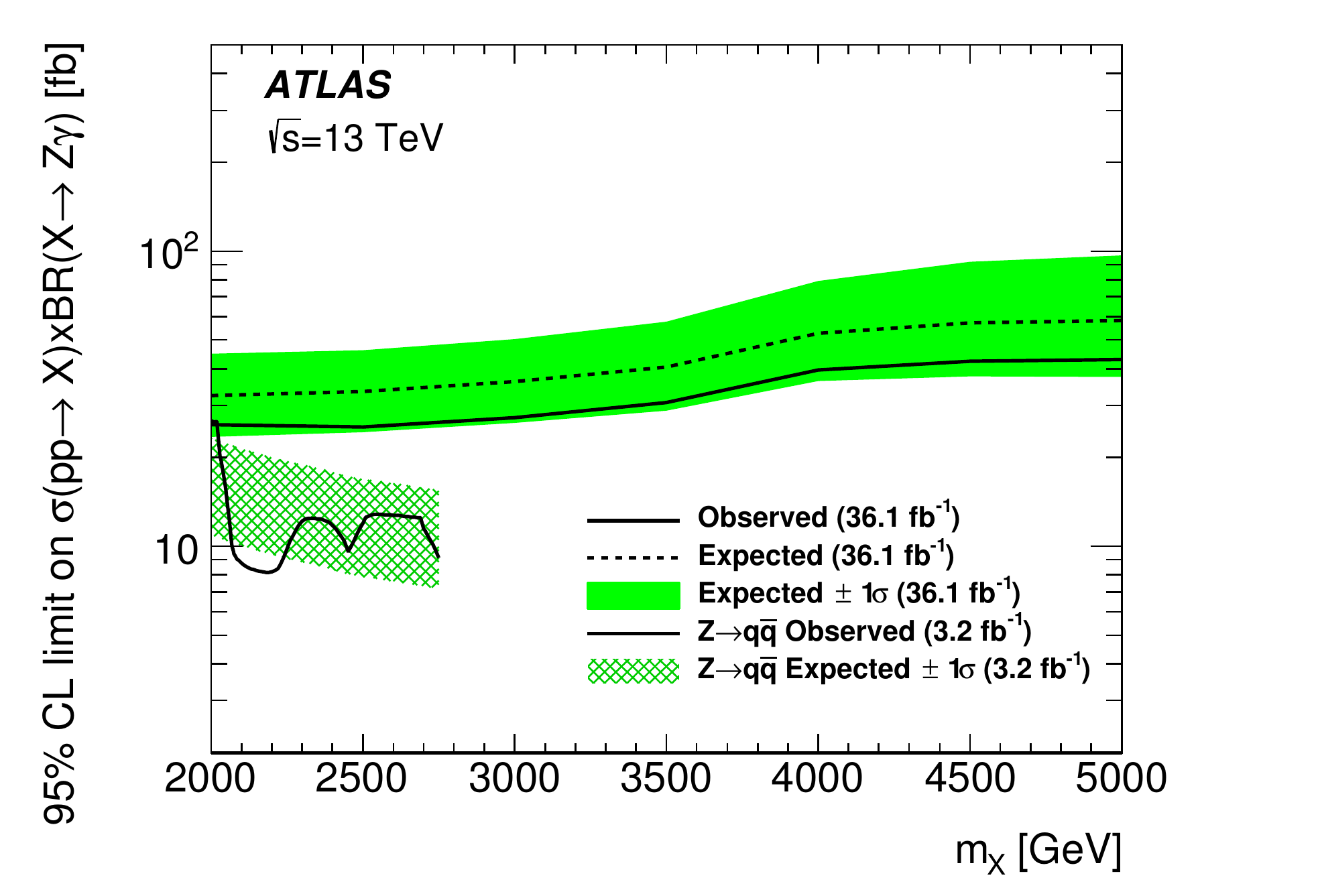}
	\caption{The observed (expected) limit at 95\% CL on the production cross section of a $Z\gamma$ resonance 
as a function of  its mass.	The limits from the search in the $Z \rightarrow q\bar{q}$ channel with 3.2 \ifb~\cite{Zgamma} are also reported.
	 }
   \label{fig:Zgamma}
\end{figure}

%% file: Conclusion.tex
Results are reported from a search for dark matter in events with a high transverse energy photon and large missing transverse momentum in $pp$ 
collisions at $\sqrt{s}=13\TeV$ at the LHC. Data collected by the ATLAS experiment and corresponding to an integrated luminosity of 36.1 $\ifb$ are used.
 The observed data are consistent with the Standard Model expectations.
 The observed (expected) upper limits on the fiducial cross section for the production of events with a photon and large 
  missing transverse momentum are 7.0 and 2.3 fb (10.6 and 3.0 fb) at 95\% CL for $\met$ thresholds of $150\GeV$ and $300\GeV$, respectively.
 For the simplified dark-matter model considered, the search excludes axial-vector and vector mediators with 
masses below  $750$--$1200\GeV$ for  $\chi$ masses below $230$--$480\GeV$ at 95\% CL, depending on the couplings chosen.
For an EFT $\gamma\gamma\chi\bar{\chi}$ model of dark-matter production, values of  the suppression scale $M_*$ up to $790\GeV$ 
are excluded at 95\% CL and the effect of truncation for various coupling values is reported.
The observed (expected) limits at 95\% CL on the production cross section for a  narrow $Z\gamma$  scalar resonance are 
26 and 43  fb (32 and 58 fb) for masses of 2 and $5\TeV$, respectively.

%% file: Acknowledgements.tex

We thank CERN for the very successful operation of the LHC, as well as the
support staff from our institutions without whom ATLAS could not be
operated efficiently.

We acknowledge the support of ANPCyT, Argentina; YerPhI, Armenia; ARC, Australia; BMWFW and FWF, Austria; ANAS, Azerbaijan; SSTC, Belarus; CNPq and FAPESP, Brazil; NSERC, NRC and CFI, Canada; CERN; CONICYT, Chile; CAS, MOST and NSFC, China; COLCIENCIAS, Colombia; MSMT CR, MPO CR and VSC CR, Czech Republic; DNRF and DNSRC, Denmark; IN2P3-CNRS, CEA-DSM/IRFU, France; SRNSF, Georgia; BMBF, HGF, and MPG, Germany; GSRT, Greece; RGC, Hong Kong SAR, China; ISF, I-CORE and Benoziyo Center, Israel; INFN, Italy; MEXT and JSPS, Japan; CNRST, Morocco; NWO, Netherlands; RCN, Norway; MNiSW and NCN, Poland; FCT, Portugal; MNE/IFA, Romania; MES of Russia and NRC KI, Russian Federation; JINR; MESTD, Serbia; MSSR, Slovakia; ARRS and MIZ\v{S}, Slovenia; DST/NRF, South Africa; MINECO, Spain; SRC and Wallenberg Foundation, Sweden; SERI, SNSF and Cantons of Bern and Geneva, Switzerland; MOST, Taiwan; TAEK, Turkey; STFC, United Kingdom; DOE and NSF, United States of America. In addition, individual groups and members have received support from BCKDF, the Canada Council, CANARIE, CRC, Compute Canada, FQRNT, and the Ontario Innovation Trust, Canada; EPLANET, ERC, ERDF, FP7, Horizon 2020 and Marie Sk{\l}odowska-Curie Actions, European Union; Investissements d'Avenir Labex and Idex, ANR, R{\'e}gion Auvergne and Fondation Partager le Savoir, France; DFG and AvH Foundation, Germany; Herakleitos, Thales and Aristeia programmes co-financed by EU-ESF and the Greek NSRF; BSF, GIF and Minerva, Israel; BRF, Norway; CERCA Programme Generalitat de Catalunya, Generalitat Valenciana, Spain; the Royal Society and Leverhulme Trust, United Kingdom.

The crucial computing support from all WLCG partners is acknowledged gratefully, in particular from CERN, the ATLAS Tier-1 facilities at TRIUMF (Canada), NDGF (Denmark, Norway, Sweden), CC-IN2P3 (France), KIT/GridKA (Germany), INFN-CNAF (Italy), NL-T1 (Netherlands), PIC (Spain), ASGC (Taiwan), RAL (UK) and BNL (USA), the Tier-2 facilities worldwide and large non-WLCG resource providers. Major contributors of computing resources are listed in Ref.~\cite{ATL-GEN-PUB-2016-002}.

%% file: atlas_authlist.tex
\begin{flushleft}
{\Large The ATLAS Collaboration}

\bigskip

M.~Aaboud$^\textrm{\scriptsize 137d}$,
G.~Aad$^\textrm{\scriptsize 88}$,
B.~Abbott$^\textrm{\scriptsize 115}$,
J.~Abdallah$^\textrm{\scriptsize 8}$,
O.~Abdinov$^\textrm{\scriptsize 12}$$^{,*}$,
B.~Abeloos$^\textrm{\scriptsize 119}$,
S.H.~Abidi$^\textrm{\scriptsize 161}$,
O.S.~AbouZeid$^\textrm{\scriptsize 139}$,
N.L.~Abraham$^\textrm{\scriptsize 151}$,
H.~Abramowicz$^\textrm{\scriptsize 155}$
% ...
\bigskip
\\
% ...
$^{8}$ Department of Physics, The University of Texas at Arlington, Arlington TX, United States of America\\
% ...
$^{12}$ Institute of Physics, Azerbaijan Academy of Sciences, Baku, Azerbaijan\\
% ...
$^{88}$ CPPM, Aix-Marseille Universit{\'e} and CNRS/IN2P3, Marseille, France\\
% ...
$^{115}$ Homer L. Dodge Department of Physics and Astronomy, University of Oklahoma, Norman OK, United States of America\\
% ...
$^{119}$ LAL, Univ. Paris-Sud, CNRS/IN2P3, Universit{\'e} Paris-Saclay, Orsay, France\\
% ...
$^{137}$ $^{(a)}$ Facult{\'e} des Sciences Ain Chock, R{\'e}seau Universitaire de Physique des Hautes Energies - Universit{\'e} Hassan II, Casablanca; $^{(b)}$ Centre National de l'Energie des Sciences Techniques Nucleaires, Rabat; $^{(c)}$ Facult{\'e} des Sciences Semlalia, Universit{\'e} Cadi Ayyad, LPHEA-Marrakech; $^{(d)}$ Facult{\'e} des Sciences, Universit{\'e} Mohamed Premier and LPTPM, Oujda; $^{(e)}$ Facult{\'e} des sciences, Universit{\'e} Mohammed V, Rabat, Morocco\\
% ...
$^{139}$ Santa Cruz Institute for Particle Physics, University of California Santa Cruz, Santa Cruz CA, United States of America\\
% ...
$^{151}$ Department of Physics and Astronomy, University of Sussex, Brighton, United Kingdom\\
% ...
$^{155}$ Raymond and Beverly Sackler School of Physics and Astronomy, Tel Aviv University, Tel Aviv, Israel\\
% ...
$^{161}$ Department of Physics, University of Toronto, Toronto ON, Canada\\
% ...
$^{a}$ Also at Department of Physics, King's College London, London, United Kingdom\\
$^{b}$ Also at Institute of Physics, Azerbaijan Academy of Sciences, Baku, Azerbaijan\\
$^{c}$ Also at Novosibirsk State University, Novosibirsk, Russia\\
$^{d}$ Also at TRIUMF, Vancouver BC, Canada\\
$^{e}$ Also at Department of Physics {\&} Astronomy, University of Louisville, Louisville, KY, United States of America\\
% ...
$^{*}$ Deceased
\end{flushleft}
